\documentclass[10pt]{article}
\usepackage[margin=1in]{geometry}
\usepackage{amsfonts,amsmath,amssymb}
\usepackage[none]{hyphenat}
\usepackage{fancyhdr}
\usepackage{graphicx}
\usepackage{float}
\usepackage[nottoc,notlot,notlof]{tocbibind}
\usepackage{bigints}

\usepackage{amssymb,amsmath,tikz}
\usepackage{pgfplots} 
\usepackage{amsthm}
\usepackage{mathrsfs}
\usepackage{pifont}
\usepackage{slashed}
\usepackage{mathtools}
\usepackage{cancel}

\showoutput
\DeclareMathOperator*{\p}{p}

\DeclareMathOperator*{\q}{k}

\newcommand*{\ud}{\mathrm{\,d}}

\theoremstyle{plain}

\newtheorem*{twr*}{Theorem}
\newtheorem*{lem*}{Lemma}
\newtheorem{twr}{Theorem}

\newtheorem*{defin*}{Definition}

\newtheorem*{rem*}{Remark}

\newtheorem{cor*}{Corollary}

\newtheorem*{notn*}{Notation}
\newtheorem*{wiener-ito*}{Wiener-It\^o-Segal Decomposition}
\newtheorem*{prop*}{Proposition}

\DeclareMathAlphabet{\mathpzc}{OT1}{pzc}{m}{it}

\begin{document}

\begin{titlepage}
\begin{center}
\vspace*{1cm}
\large{\textbf{CAUSAL PERTURBATIVE QED AND WHITE NOISE}}\\
\vspace*{2cm}
\small{JAROS{\L}AW WAWRZYCKI}\\[1mm]
\tiny{Bogoliubov Labolatory of Theoretical Physics}
\\
\tiny{Joint Institute of Nuclear Research, 141980 Dubna, Russia}
\\
\tiny{jaroslaw.wawrzycki@wp.pl}
\\
\vspace*{1cm}
\tiny{\today}\\
\vfill
\begin{abstract}
We present the Bogoliubov's causal perturbative QFT, which includes 
only one refinement: the creation-annihilation operators at a point, \emph{i.e.} 
for a specific momentum, are mathematically 
interpreted as the Hida operators from the white noise analysis. 
We leave the rest of the theory completely unchanged. This allows avoiding 
infrared-- and ultraviolet -- divergences in the transition 
to the adiabatic limit for interacting fields and elimination of the 
free parameters of the theory, associated with the choice of normalization in computation of the
retarded and advanced parts of causal distributions (corresponding to the freedom 
in the choice of renormalization scheme). This enhances the predictive power of the theory, and in particular 
allows to derive non-trivial mass relations.
The approach is general and can be applied to investigate any perturbative QFT.    
\end{abstract}
\vspace*{0.5cm}
\tiny{{\bf Keywords}: scattering operator, causal perturbative method in QFT, \\
white noise, Hida operators, integral kernel operators, Fock expansion}
\end{center}
\vfill
\end{titlepage}



\section{Introduction}

It was Bogoliubov \cite{Bogoliubov_Shirkov} who recognized the fundamental role of causality principle
in perturbative QFT, which allowed afterward  the rigorous and axiomatic treatment of the perturbative 
method and clarified the renormalization prescription in QFT. Let us remind his idea and a subsequent development. 
It is based on the scattering
generalized operator $S$.
Suppose we are given an interaction Lagrangian $g_{{}_{0}} \mathcal{L}_{{}_{0}}(x)$ with a coupling
constant $g_{{}_{0}}$. In order to implement and then make use of the causal relations (inspired by Schwinger's
treatment of the external field problem) $g_{{}_{0}}$ is replaced in \cite{Bogoliubov_Shirkov} by a smooth
function on the space-time, plying a role analogous to a classical field,  say, ``intensity of interaction'',
or ``switching off'' test function $g_{{}_{0}}$. In fact,
the method is quite general and, after Schwinger and \cite{Bogoliubov_Shirkov}, we consider a generalized interaction
Lagrangian, into which we introduce additional terms multiplied by their own ``switching off'' test functions $g_{{}_{j}}$.  
We consider a generalized interaction Lagrangian with multicomponent switching off test function 
$g = (g_{{}_{0}}, \ldots, g_{{}_{k}})$
\begin{equation}\label{L}
\mathcal{L}(x) = \sum\limits_{j=0}^{k} g_{{}_{j}}(x) \mathcal{L}_{{}_{j}}(x) = g_{{}_{0}}(x) \mathcal{L}_{{}_{0}}(x) +
\sum\limits_{j=1}^{k} g_{{}_{j}}(x) W_{{}_{j}}(x), 
\end{equation}
and with $\mathcal{L}_j(x)$ equal to any Wick products $W_{{}_{j}}(x)$ of free fields, in most cases of even degree in Fermi fields,
and with $\mathcal{L}_{{}_{0}}$ being the true interaction
Lagrangian, with $g_{{}_{0}}$ then eventually tending to the physical coupling constant 
(adiabatic limit problem), and with the other terms in $\mathcal{L}$ introduced after Schwinger
in order to compute the interacting counterparts $W_{{}_{j \, \textrm{int}}}$ of the free Wick products 
$W_{{}_{j}}$, $j>0$, or for the treatment of the external field problem. In case the Wick monomial 
$W_{{}_{j}}(x)$ is odd in Fermi fields, the corresponding $g_{{}_{j}}$ is a test function with values in the Grassmann 
algebra with involution in the sense of \cite{Berezin}. Nontrivial restrictions on the allowed interaction term 
$\mathcal{L}_{{}_{0}}(x) = W_{{}_{0}}(x)$ will come e.g. from the renormalizability of the perturbation quantum field theory, 
and further from the concrete mathematical refinement of the axioms initiated in  \cite{Bogoliubov_Shirkov}, 
which we present below. Specification of the free fields and their Wick products as kinds of generalized operators (functionals of test functions)  
and the class of allowed test functions $g$ in (\ref{L}) will be specified later, and of course lies at the very heart of the problem.

To the interaction Lagrangian (\ref{L}) there corresponds the generalized scattering operator $S(g)$ and its inverse $S(g)^{-1}$, 
which becomes a functional  of the switching off test function $g$, and which is postulated in \cite{Bogoliubov_Shirkov} 
as a formal power series in $g$:  
\begin{align}
S(g) = \boldsymbol{1} + \sum\limits_{n=1}^{\infty} {\textstyle\frac{1}{n!}} S_n(g^{\otimes \, n}),
\,\,\,\,\,\,
S(g)^{-1} = \boldsymbol{1} + \sum\limits_{n=1}^{\infty} {\textstyle\frac{1}{n!}} \overline{S_n}(g^{\otimes \, n}),
\label{PowerSeriesS}
\\
S_n(g^{\otimes \, n}) = 
\sum\limits_{j_1,\ldots,j_n=0}^{k}\bigintsss \ud^4 x_1 \ldots \ud^4x_n \, S_n(j_1,x_1, \ldots, j_n, x_n) \, g_{{}_{j_1}}(x_1) \ldots g_{{}_{j_n}}(x_n),
\label{Sn(x1,...,xn)}
\end{align}
and similarly for $\overline{S_n}(g^{\otimes \, n})$.  It was established in \cite{Bogoliubov_Shirkov} that $S(g)$
respects the axioms  of  (I) causality: $S(g+h)= S(g)S(h)$ whenever there exists a Lorentz frame in which 
the support of $h$ lies before the support of $g$, (II) unitarity: $S(g)^{-1}= S(g)^{+}$ 
(or Krein isometricity: $S(g)^{-1}= \eta S(g)^{+}\eta$ if gauge fields are present,
with $\eta$ being the Gupta-Bleuler operator),
(III) relativistic covariance and (IV) correspondence principle: $S_1(g) = \int \mathcal{L}(x) \ud^4 x$, where  $\mathcal{L}(x)$
is given by (\ref{L}).

Having given $S(g)$ and its inverse $S(g)^{-1}$, the local interacting fields $W_{{}_{j \,\, \textrm{int}}}(g_{{}_{0}},\phi)$
corresponding to the free Wick monomials $W_{{}_{j}}(\phi)$, evaluated at the space-time test function $\phi$, 
are constructed as the formal variational derivatives \cite{Bogoliubov_Shirkov}:
\[
W_{{}_{j \,\, \textrm{int}}}(g_{{}_{0}},\phi) = \bigintsss \left[
S(g)^{-1}
{\textstyle\frac{i\delta S(g)}{\delta g_{{}_{j}}(x)}} 
\right]\Bigg|_{{}_{g_{{}_{i\neq0}}=0}} \phi(x) dx.
\]

The Bogoliubov axioms (I)-(IV) can be expressed in terms of the kernels $S_n(j_1,x_1, \ldots, j_n, x_n)$
of the integrals (\ref{Sn(x1,...,xn)}) in the following manner 
\begin{enumerate}
\item[(I)]
\begin{gather*}
S_n(j_1,x_1, \ldots, j_n,x_n) = (-1)^{s(X,Y)} \, S_k(j_{r_1},x_{r_1}, \ldots, j_{r_k},x_{r_k})S_{n-k}(j_{r_{k+1}},x_{r_{k+1}}, \ldots, j_{r_n}, x_{r_n}),
\\
\,\,\,\, \textrm{whenever $\{j_{r_{k+1}},x_{r_{k+1}}, \ldots, j_{r_{n}}, x_{r_n} \} \preceq \{j_{r_1},x_{r_1}, \ldots, j_k,x_{r_k}\}$}.  
\end{gather*}
\item[(II)]
\[
U_{b,\Lambda} S_n(j_1,x_1, ..,j_n,x_n) U_{b, \Lambda}^{+} = 
\sum\limits_{j'_1, \ldots, j'_n} 
V_{{}_{j_1 \,\, j'_{1}}} \cdots V_{{}_{j_n \,\, j'_{n}}}S_n(j'_1,\Lambda^{-1}x_1 - b, ..,j'_n, \Lambda^{-1}x_n - b),
\]
\item[(III)]
\[     
\sum\limits_{j_1, \ldots, j_n} \overline{S}_{n}(j_1,x_1, \ldots, j_n,x_{n}) = \sum\limits_{j_1, \ldots, j_n}  \eta S_n(j_1,x_1, \ldots, j_n,x_{n})^{+} \eta,
\]
\item[(IV)]
\[                  
S_{1}(j,x) = i \mathcal{L}_{{}_{j}}(x), 
\]
where $\mathcal{L}_{{}_{j}}(x)$ is the interaction Lagrangian density operator in (\ref{L}). 
For each $k$, the index $j_k$ has the range of the index $j$ in (\ref{L}).
\end{enumerate}
In order to explain the notation, let $Z$ denote the set $\{j_1,x_1, \ldots, j_n,x_n\}$ of variables. 
Here we have a partition $Z= X \sqcup Y$ of $Z$ into two disjoint subsets.
In the partition, we treat each pair $j_i,x_i$ as a single element. $X= \{j_{r_1},x_{r_1}, \ldots, j_{r_k},x_{r_k} \}$,
$Y = \{j_{r_{k+1}},x_{r_{k+1}}, \ldots, j_{r_n}, x_{r_n}\}$. Symbol $s(X,Y)$ denotes the parity of permutation
of Grassmann variabes in the permutation $Z \rightarrow (X,Y)$.
In (II) we have the matrices $V_{{}_{j \, j'}}$ coming from the transformation formulas 
$U_{{}_{b,\Lambda}}W_{{}_{j}}(x)U^{+}_{{}_{b,\Lambda}} = \underset{j'}{\Sigma}V_{{}_{j \, j'}}W_{{}_{j'}}(\Lambda^{-1}x-b)$ 
of the Wick monomials $W_{{}_{j}}$ in  (\ref{L}). The kernels $S_n(j_1,x_1, \ldots, j_n, x_n)$  have the general form of linear combinations
of the Wick products of free fields with coefficients equal to translationally invariant tempered distributions (``Green functions'').

Rigorous definition of the class of generalized operators including $S_n$ is not specified in \cite{Bogoliubov_Shirkov}.    
It is only remarked in \cite{Bogoliubov_Shirkov} that this class should include Wick products of free fields with coefficients 
equal to translationally invariant tempered distributions. It was recognized in \cite{Bogoliubov_Shirkov}, \S 29, that having given the class 
specified in whatever rigorous manner, which allows as test functions $g$ the Schwartz test functions, the axioms (I)-(IV) determine 
the kernels $S_n(j_1,x_1, \ldots, j_n, x_n)$  of all orders up to a generalized operators $\Lambda_{n}(j_1,x_1, \ldots, j_n, x_n)$
supported at the full diagonal. As proved in \cite{Bogoliubov_Shirkov}, this ambiguity is precisely the ambiguity which
corresponds to the ordinary ambiguity in the renormalization prescription. However, in \cite{Bogoliubov_Shirkov}, \S 29.2, it is only outlined
the existence proof for $S_n$ (with the remarked ambiguity). In passing to the construction of $S_n$ the rigorous approach, indicated in \S 29.2,
is abandoned. Instead, it was observed in \cite{Bogoliubov_Shirkov} that from (I)-(IV) it follows that, outside the full diagonal 
($x_i\neq x_k$ for some $i\neq k$), $S_n(j_1,x_1, \ldots, j_n, x_n)$ is equal to the ordinary chronological product 
$T\left[\mathcal{L}_{{}_{j_1}}(x_1) \cdots \mathcal{L}_{{}_{j_n}}(x_n)\right]$ using ordinary multiplication by the step 
theta function, and this formula is formally regarded as if it was true in the whole domain of the space-time variables. This formal
extension leads to divergent terms, but the divergent part has precisely the full diagonal form $\Lambda_n$ mentioned above,
with divergent coefficients, which (in case of renormalizable $\mathcal{L}_{{}_{0}}$) can be subsumed by addition of finite number of terms in the Lagrangian
of the same kind as the original Lagrangian, but with infinite coefficients. Thus, infinities can be eliminated, by addition of a 
finite number of counterterms to the Lagrangian, with infinite coefficients (renormalization prescription). 
But the rigorous existence proof \cite{Bogoliubov_Shirkov}, outlined in \S 29.2, suggests that in principle it should also 
be possible to \emph{construct} or \emph{compute} $S_n$ (with the mentioned ambiguity)
in a rigorous manner, without resorting to such infinite subtractions. Such rigorous \emph{construction}, based on (I)-(IV), was indeed 
given later by Epstein and Glaser \cite{Epstein-Glaser} for the scalar massive field (subsequently applied to other Lagrangians, 
including QED, by other authors \cite{DKS1}-\cite{DKS4}). It is obvious that in order to give concrete and rigorous mathematical content to the axioms (I)-(IV),  
we need to specify the class of generalized operators to which $S_n$ belong, as well as the class of space-time test functions. 
Otherwise, no rigorous construction or calculation of $S_n$ based on the axioms (I)-(IV) would be possible.     
Epstein and Glaser \cite{Epstein-Glaser} assumed that the free fields, their Wick products, (tensor) products $W(x)W(y)$
of the Wick products of free fields, and $S_n(j_1,x_1, \ldots, j_n, x_n)$ are generalized operators in the sense 
of operator valued distributions defined by Wightman \cite{wig}. For this class of generalized operators, the class of test
functions $g$ includes the Schwartz rapidly decreasing functions. Having the content of the axioms (I)-(IV) established in this manner,
they were able to construct inductively $S_n$. In fact the rigorous inductive construction \cite{Epstein-Glaser} of $S_n$ out of $S_k$, $k \leq n-1$, 
is based on the observations already made in \cite{Bogoliubov_Shirkov}. Let, for simplicity of notation, each pair of variables 
$j_k,x_k$ in the kernel $S_n$ be shortly written by $x_k$, remembering that each $x_k$ is a variable of various possible kinds,
correspondingly to the possible values of the index $j_k$ coinciding with the range $\{0,\ldots, k\}$ of the index $j$ in (\ref{L}):
in particular, $x_k$ is of Grassmann type if the corresponding $g_{{}_{j_k}}$ is Grassmann valued. 
Let, further, $Z$ be the set of variables $\{x_1, \ldots, x_{n-1}\}$. Consider the sums $A'_{n}(Z,x_n) = \Sigma (-1)^{s(X,Y,x_n)}\overline{S}(X)S(Y,x_n)$
and $R'_{n+1}(Z,x_n) = \Sigma (-1)^{s(Y,x_n,X)} S(Y,x_n) \overline{S}(X)$ over all divisions 
$Z = X \sqcup Y$ of $Z$ into two disjoint sets, excluding $X = \emptyset$.     
Here $s(X,Y,x_n)$ is the sign of the permutation of the Grassmann-type variables in passing from the order $(Z,x_n)$ to the order $(X,Y,x_n)$
and the subscript $k$ at $S_k(X)$ has been omitted, being equal to the number of elements of $X$.
Next, consider $A_{n}(Z,x_n) = \Sigma (-1)^{s(X,Y,x_n)}\overline{S}(X)S(Y,x_n) = A_{n}(Z,x_n) + S_n(Z,x_n)$
and $R_{n}(Z,x_n) = \Sigma (-1)^{s(Y,x_n,X)} S(Y,x_n) \overline{S}(X) = R_{n}(Z,x_n) + S_n(Z,x_n)$ with summation 
over all divisions of $Z$ including $X = \emptyset$.  Thus $A'_n,R'_n$ can be computed from $S_k$, $k\leq n-1$. Then, as already observed in
\cite{Bogoliubov_Shirkov}, \S 21.2 formula (13), $A_{n}(Z,x_n)$ and $R_{n}(Z,x_n)$, have, respectively, advanced and retarded
supports, restricted, respectively, to the set of  $x_k$, $k=1, \ldots, n-1$, each of which lies in the past light cone emerging from $x_n$
or, respectively, in the forward light cone emerging from $x_n$. The point is that the said support properties of $A_n$ and $R_n$ follow 
from the axioms (I)-(IV) and allow computation of $S_n$ in the following manner.  Because $D_n = R'_n - A'_n = R_n-A_n$, then $D_n$
has causal support, with $A_n$ being the advanced and $R_n$ the retarded part of $D_n$,
with $D_n$ which can be computed out of $S_k$, $k \leq n-1$. This means that all the coefficient tempered distributions in the Wick decomposition
of $D_n$ have causal support. Because moreover they have finite singularity order $\omega$ at zero, they can be splitted into retarded and advanced part,
up to the finite linear combinations of the Dirac delta and its derivatives -- distributions supported at the full diagonal, \emph{i.e.} up to the
freedom depending on a finite number of constants depending, in turn, on the singularity order $\omega$ of the splitted causal distribution. Singularity
order at zero (in space-time coordinates) is undestood here in the standard sense \cite{Vladimirov1}, and coincides 
(for the Fourer transformed distribution)
with the degree of the polynomial growth at infinity in momentum space (for function-like Fourier transform of plynomial growth), 
and coincides with the ordinary divergence degree of the corresponding graph in the momentum space \cite{Bogoliubov_Shirkov}. 
Thus, the retarded and advanced  parts $R_n,A_n$ of $D_n$ can be computed independently of (I)-(IV) on using the splitting theory 
of causal distributions having finite singularity order. Therefore, we can compute $S_n = \textrm{adv} D_n - A'_n = \textrm{ret} D_n - R'_n$. 
Strictly speaking, we have used one implicit assumption here: that 
\begin{enumerate}
\item[(V)]
The advanced and retarded parts of the splitted causal distribution $d_n$
have the same singularity order as $d_n$, 
\end{enumerate}
which should be added to the axioms (I)-(IV), in order to base the whole computation of $S_n$
solely on  (I)-(V), without any additional implicit assumptions, remembering also that in addition we have interpreted the free
fields, their Wick products, and higher order contributions $S_n$, as the generalized operators in the Wightman sense \cite{wig}.
$S_n,R_n,A_n$ are, respectively, called time ordered products, retarded products, and advanced products.

In this manner, using (I)-(V) and said interpretation of generalized operators, Epstein and Glaser \cite{Epstein-Glaser}  
computed $S_n$ rigorously, without any reference to
infinite subtractions, eliminating all ultraviolet infinities, in the computation of the kernels $S_n$ with the ambiguity in $S_n$
coming from the finite non-negative singularity order $\omega$ of the coefficients of $D_n$. Theory is renormalizable 
if the singularity order $\omega$ of each contribution to $D_n$,
equal to the singularity order of each corresponding contribution to  $S_n$, 
is bounded by a constant independent of $n$ and equal $4$ minus the number of external lines, counted with a weight depending on the spin
of the external line, and minus the number of derivatives in external lines. This puts nontrivial, well-known,
restrictions on $\mathcal{L}_{{}_{0}}$ in (\ref{L}). \emph{I.e.} in renormalizable case, the singularity order $\omega$ of a term in $S_n$
 corresponding to a set of external lines  (\emph{i.e.} of the term proportional to the Wick product of free 
fields represented by these external lines) is less than or equal $4$ minus 
a positive number depending on the number of these external lines and derivatives in these external lines.  
E.g. for the spinor QED Lagrangian $\mathcal{L}_{{}_{0}}$, $\omega \leq 4 - (3/2\mathfrak{f}+\mathfrak{k})$, 
with $\mathfrak{f}$ equal to the total number of fermion external lines and $\mathfrak{k}$ equal to the total number of external photon lines. 
For the Yang-Mills Lagrangian (without couplings to matter fields) \cite{DKS3}, $\omega \leq 4-\mathfrak{b}-\mathfrak{g}-\overline{\mathfrak{g}}-\mathfrak{d}$, 
where $\mathfrak{b}$ is the number of external gluon lines, $\mathfrak{g}, \overline{\mathfrak{g}}$ -- external ghost, antighost lines, 
and $\mathfrak{d}$-the number of derivations in the external gluon and antighost lines. Therefore, the set of all contributions
which require renormalization, \emph{i.e.} those with $\omega \geq 0$, correspond to a finite number of different sets 
of external lines (Wick products of free fields), and theory can be renormalized. 
The scattering operator, based on the axioms (I)-(V)
specified as in \cite{Epstein-Glaser}, gives the coefficient distributions of $S_n$ exactly the same
as the (finite) renormalized coefficients obtained with the method using renormalization \cite{Bogoliubov_Shirkov}. Only the source
of the ambiguity in $S_n$ is differently looked at: in the rigorous approach \cite{Epstein-Glaser} this ambiguity follows from the existence
of the coefficients in $D_n$ with non-negative singularity order $\omega$ implying non-unique splitting. 
In the approach using renormalization it comes from the fact that 
there is no distinguished finite value which could represent the difference of two positive and infinite numbers 
(\emph{i.e.} the symbol $\infty-\infty$ is indefinite). Below we return to a deeper interpretation of this ambiguity.

The reader may therefore ask: what is the profit of this rigorous formulation of the axioms (I)-(V) and calculation of $S_n$
based on (I)-(V)? A possible answer is this\footnote{One important motivation is the clear separation of the Ultra-Violet (UV)- and Infra-Red (IR)-divergence problems: the UV-infinities are suited in the splitting of causal distributions into retarded and advanced parts, and the IR-infinities are located in the
adiabatic limit $g_{{}_{0}}\rightarrow const.$ problem. Another benefit is a significant simplification of the analysis 
of the renormalizability and unitarity of the theory with non-abelian gauges \cite{DKS1}-\cite{DKS4}, \cite{Grigore}.}: 
by making a theory rigorous, we expect to strengthen its predictive power, simply reducing 
its statements to the logical consequences of the axioms. In our case, for example, the ambiguity in the splitting (ambiguity in renormalization scheme 
in the approach using renormalization) can be eliminated (or substantially reduced) 
by imposing existence of the adiabatic limit $g_{{}_{0}} \rightarrow constant$ for the scattering operator 
or for the interacting fields. For this reason, contribution of  \cite{Epstein-Glaser}, making
the theory rigorous, gives hope to this perspective. However, the above Bogoliubov-Epstein-Glaser formulation, 
based on the said mathematical interpretation of the generalized operators, and axioms (I)-(V), has important shortcomings.
Namely, in the most interesting cases of QFT with infinite range of interaction, as e.g. QED, the adiabatic limit
for interacting fields does not exist. In fact this could have been expected, as the Wightman operator distributions have, by construction,
the Schwartz functions as their test functions, and the constant function is not the element of the Schwartz test function space,
with the further specifications needed to fix the sense of the adiabatic limit, and with nontrivial existence problems.
Therefore, we expect existence of this limit only in some exceptional theories (e.g. massive scalar field \cite{epstein-glaser-al}).

In order to free the theory of Bogoliubov-Epstein-Glaser based on (I)-(V) from the said shortcomings, 
we no longer regard the generalized operators which include the free fields and their Wick products with coefficients 
equal to translationally invariant tempered distributions, as the generalized operators in the Wightman sense. 
In fact, the free fields, their Wick products with coefficients equal to
any translationally invariant tempered distributions, can also be regarded as particular cases of the finite linear combinations of the
so-called integral kernel generalized operators with vector valued kernels of the white noise calculus in the Fock space \cite{obataJFA}.
Therefore, in quantum field theory, one can actually consider free fields, their Wick products, and higher-order contributions
$S_n$ as finite sums of generalized integral kernel operators of the white noise calculus \cite{WN}. 
This allows us to go much further in understanding the adiabatic limit problem than was possible in the approach based on the
generalized Wightman operators. In this paper, we will focus ourselves on the application of \cite{WN} to QED, 
where we have the existence of the adiabatic limit for interacting fields as generalized integral kernel operators with vector valued kernels
if and only if the charged particle has nonzero mass (Theorems \ref{electronmass=/=0}, \ref{electronmass=0}, Section \ref{QEDapplication} proved in \cite{IF}), 
and where we have the existence of the natural product operation for the interacting fields in the adiabatic limit, whenever it exists 
(Theorem \ref{product}, Section \ref{QEDapplication}), and finally, where we have the existence of the 
adiabatic limit for the scattering operator understood as a generalized integral kernel operator (Section \ref{QEDapplication}). 
Next, we give some further perspectives, including other interactions (Section \ref{Perspectives}). 
Finally, we give a comparison with other approaches (Section \ref{Alternatives}). 
In Section \ref{Hida} we remind the main idea of
\cite{WN}, recall some basic facts concerning generalized integral kernel operators with vector valued kernels, explain the way in which
we apply them to the Bogoliubov-Epstein-Glaser perturbative QFT, and recall the definition of Hida operators.

\section{Axioms for $S$ with Hida operators. Hida operators}\label{Hida}

Therefore, we interpret the free fields, the Wick products, and the operators $S_n$ as the 
integral kernel operators with vector valued kernels of the white noise calculus \cite{obataJFA}.
In fact, this means that we regard the creation-annihilation operators at specific spin-momenta $\boldsymbol{\p}$ as the
Hida operators $\partial_{\boldsymbol{\p}}^{*}, \partial_{\boldsymbol{\p}}$ of the white noise calculus \cite{obataJFA}, which indeed respect the
canonical commutation rules. Except for this choice of interpretation of the creation-annihilation operators, we leave the theory,
subsumed in (I)-(V), completely unchanged. The class of generalized operators to which $S_n$ belong becomes now 
substantially extended, so that the adiabatic limit $g_{{}_{0}} \rightarrow constant$ exists for $n$-order contributions
to interacting fields for realistic QFT with the normalization in the splitting uniquely determined by the condition requiring 
existence of this limit. Equivalently, we keep the theory (I)-(V), together with the inductive step based on the splitting of causal distribution,
but regard each free field $\mathbb{A}(x)$ as the sum of two integral kernel operators with vector valued kernels $\kappa_{0,1}, \kappa_{1,0}$ 
in the sense \cite{obataJFA}:
\begin{equation}\label{FreeField=Xi(xi01)+Xi(xi10)}
\mathbb{A}(x) = \int  \kappa_{0,1}(\boldsymbol{p};x) \partial_{\boldsymbol{\p}} \, \ud^3\boldsymbol{p}
+ \int \kappa_{1,0}(\boldsymbol{p};x) \partial_{\boldsymbol{\p}}^{+} \, \ud^3\boldsymbol{p},
\end{equation}
first one corresponding to the negative and second one to the positive energy part (with the only change of the convention in comparison to the
one used in mathematical literature, that our $\partial_{\boldsymbol{\p}}^{+}$ is the linear transpose $\partial_{\boldsymbol{\p}}^{*}$
preceded and followed by the complex conjugation instead of being simply the linear transpose), with $\kappa_{0,1}, \kappa_{1,0}$ 
being the ordinary negative and positive energy plane wave solutions of the linear hyperbolic equation associated to the free field
$\mathbb{A}$, which we may regard as (function-like) distribution of the spin-momenta variables $\boldsymbol{\p}$ and space-time
variables $x$. We know that each free field is associated with the fixed orbit of a fixed point in momentum space under the natural action
of the Lorentz group (positive energy hyperboloid of mass $m$ correspondingly to the mass $m$ of the field $\mathbb{A}$, degenerating to the
positive energy cone without the apex in case $m=0$). The single particle Hilbert space $\mathcal{H}$ of a (scalar, fourvector, \emph{e.t.c}) field
is given by the Fourier transforms of  square summable (scalar, fourvector, \emph{e.t.c}) functions $f$ over space-time
restricted to the corresponding positive energy orbit and muliplied by the corresponding (momentum dependent) positive energy 
idempotent (projection) for essentially neutral (``real'') fields. For (``complex'') charged fields, we have the additional direct summand
of the single particle space given by the conjugation (transposed complex conjugation) of the Fourier transforms of  square summable 
(spinor, \emph{e.t.c}) functions $f$ over space-time
restricted to the corresponding negative energy orbit and muliplied by the corresponding (momentum dependent) negative energy 
idempotent (projection).  In general, the single particle spaces in momentum space have the form of bundles over the corresponding orbits, 
with elements which cannot be regarded as ordinary
equivalence classes of (scalar, spinor, fourvector, \emph{e.t.c}) functions on the corresponding orbits. For example in the Dirac spinor
field, we have to use the non-trivial idempotents or rank $2$ ($1/2$-spin). In each case, the single particle Hilbert space possess the natural
structure of a rigged Hilbert space in the sense of \cite{GelfandIV}: $E \subset \mathcal{H} \subset E^*$, where $E$ is the
nuclear space, which we obtain when we use Schwartz functions $f$ in the construction of the elements of $\mathcal{H}$
in the massive case, or Schwartz functions for which all derivatives of the Fourier transforms vanish at zero in massless case. 
The single particle test space $E$ is of the same type as the Schwartz space itself, being the standard countably Hilbert and nuclear in the sense \cite{obataJFA},
\cite{GelfandIV}, because in each case the said indempotent (if not equal to $\boldsymbol{1}$) 
is at most of polynomial growth (for higher integer spin, or even bounded for the half spin fields) and is smooth. 
The single particle rigged Hilbert space $E \subset \mathcal{H} \subset E^*$, called nowadays Gelfand triple, 
naturally arising as above from the group representation point of view, has not the so-called standard form in case we have non-trivial
 positive and negative energy idempotents (e.g. in the sigle particle space of the Dirac field). This means that $E$ and $\mathcal{H}$ does not have the
form of ( a.e. equivalence classes of) function spaces over a measure space. 
But, in each case the idempotents define a natural unitary equivalences $U$ of the single particle Gelfand triples $E \subset \mathcal{H} \subset E^*$
with their standard realizations $E \simeq \mathcal{S}(\sqcup \mathbb{R}^3; \mathbb{C}^d) = \mathcal{S}(\mathbb{R}^3; \mathbb{C}^{kd})$, 
$\mathcal{H} \simeq L^2(\sqcup \mathbb{R}^3; \mathbb{C}^d) = L^2(\mathbb{R}^3; \mathbb{C}^{kd})$, 
$E \simeq \mathcal{S}(\sqcup \mathbb{R}^3; \mathbb{C}^d)^* = \mathcal{S}(\mathbb{R}^3; \mathbb{C}^{kd})^*$ 
for the  massive case. Here $\sqcup \mathbb{R}^3$ is the disjoint sum of a number $k$ of copies of $\mathbb{R}^3$ depending on the
spin of the field with the ordinary invariant Lebesgue measure on each copy $\mathbb{R}^3$. For the massless case we have the standard realizations
$E \simeq \mathcal{S}^0(\sqcup \mathbb{R}^3; \mathbb{C}^d)$, 
$\mathcal{H} \simeq L^2(\sqcup \mathbb{R}^3; \mathbb{C}^d)$, $E^* \simeq \mathcal{S}^0(\sqcup \mathbb{R}^3; \mathbb{C}^d)^*$,
where $\mathcal{S}^0$ is the closed subspace of the Schwartz space $\mathcal{S}$ of all those functions which have all derivatives
vanishing at zero. Unitary equivalence $U$ means that $U$ is unitary in ordinary sense, with restriction to $E$ continuous in the countably
Hilbert nuclear topology on $E$ and $\mathcal{S}$ or $\mathcal{S}^0$, and thus, by duality, with continuous dual $U^*$:
$\mathcal{S}^* \rightarrow E^*$ or $\mathcal{S}^{0*} \rightarrow E^*$. Now we can give precise definition of the sense 
in which the kernels $\kappa_{0,1}, \kappa_{1,0}$ in (\ref{FreeField=Xi(xi01)+Xi(xi10)}) can be regarded as vector valued distributions. 
In the spin-momenta variables $\boldsymbol{\p}$
they are regarded as elements of $E^*$, identified with its standard realization. Concerning the space-time variable $x$ they act on the 
ordinary Schwartz space $\mathscr{E}$ of functions on the space-time. Then $\kappa_{0,1}, \kappa_{1,0}$ can be regarded
as the continuous linear maps $E \rightarrow \mathscr{E}^*$. We will use the standard notation 
$\mathscr{L}(E, \mathscr{E}^*)$ for the linear space of such maps endowed with the topology of uniform convergence on bounded sets
\cite{obataJFA}. Therefore, $\kappa_{0,1}, \kappa_{1,0}\in\mathscr{L}(E, \mathscr{E}^*)$ are $\mathscr{E}^*$-valued
distributions on the single particle test spaces $E$, where $\mathscr{E}$ is the space-time Schwartz test space of the field 
$\mathbb{A}$. The essential point lies in the possibility of the extension $(E) \subset \Gamma(\mathcal{H}) \subset (E)^*$  
of the single particle Gelfand triple
structure $E \subset \mathcal{H} \subset E^*$ over the Fock space $\Gamma(\mathcal{H})$ 
of the field  $\mathbb{A}$, which again is a Gelfand triple in the sense \cite{GelfandIV}, which gives us
(infinite dimensional) Hida test space $(E)$ in the Fock space together with its strong dual $(E)^*$. It is based on the
abstract construction of the Gelfand triple introduced in \cite{GelfandIV}, through the standard operator $A$
on $\mathcal{H}$ associated to the triple. Operator $A$ is said to be standard whenever it is self-adjoint positive with 
some negative power $A^{-r}$ being of Hilbert-Schmidt class with $\textrm{inf} \,\, \textrm{Spec} \, A >0$. 
Recall \cite{GelfandIV}, that the Gelfand triple 
$E \subset \mathcal{H} \subset E^*$  is canonically associated with -- or determined by -- a standard $A$ on $\mathcal{H}$ if and only
if $E$ is equal to the projecive limit of the Hilbert spaces equal do the closures of  $\textrm{Dom} \, A^{k}$ with respect to
the inner producs $(\cdot , \cdot)_k = (A^k \cdot , A^k\cdot )_{{}_{\mathcal{H}}}$, $k=0,1, \ldots$, 
and $E^*$ is equal to the inductive
limit of the Hilbert spaces equal to the closures of $\mathcal{H}$ with respect to
the inner producs $(\cdot , \cdot)_{-k} = (A^{-k} \cdot , A^{-k}\cdot )_{{}_{\mathcal{H}}}$. 
The single particle Gelfand triples are associated with the standard $A$ which is a finite direct sum of the three-dimensional
oscillator Hamiltonian operator, in massive case, to which we eventually add $\boldsymbol{1}$
in order to achieve $\textrm{inf} \,\, \textrm{Spec} \, A >1$. $A$ for the massless case is different \cite{WN}. 
The fundamental observation due to Hida, is that the Gelfand operator realization of the single particle Gelfand triple can be lifted
to the whole Fock space $(E) \subset \Gamma(\mathcal{H}) \subset (E)^*$, and is associated with the Fock 
lifting $\Gamma(A)$ of the operator $A$ determining the single particle Gelfand triple. This is so, because 
$\Gamma(A)$ is standard  whenever $A$ is standard with $\textrm{inf} \,\, \textrm{Spec} \, A >1$, which is the case for the
single particle Gelfand triples \cite{obataJFA}.

Having given the Hida test space $(E)$ and its strong dual $(E)^*$, let us give the definition of the
canonical Hida operators $\partial_{\boldsymbol{\p}}^{+}, \partial_{\boldsymbol{\p}}$. Each $\Phi \in (E)$
is given by a convergent (in the nuclear topology of the Hida test space $(E)$) series of $n$ particle
states $\Phi_n \in E^{\widehat{\otimes} n}$. Analogous decomposition into $n$-particle generalized states $\Phi_{n}^{*}$ 
we have for each $\Phi^* \in (E)^*$, convergent in the strong dual topology, with each $\Phi_{n}^{*} \in E^{* \, \widehat{\otimes} n}$.
Here $E^{\widehat{\otimes} n=0} = E^{*\widehat{\otimes} n=0} = \mathbb{C}$, and $\Phi_{0},\Phi_{0}^{*},  \in \mathbb{C}$ 
being the multiples of the vacuum state. Here we consider only the projective tensor products $\otimes$ of the nuclear spaces $E,E^*$
$(E), (E)^*$, and Hilbert space tensor products, if the tensored spaces are Hilbert spaces.
The symbol $\widehat{\otimes}$ means symmetrized tensor product in case of Bose field and alternated tensor product in the Fermi case.
For each $w \in E^* \supset E$, we can define the following annihilation $a(w)$ and creation $a(w)^+$ operators
by defining component-wise their action on the arbitrary state $\Phi \in (E)$: 
\[
a(w) \Phi_0 = 0, \,\,\, a(w) \Phi_n = n \overline{w} \widehat{\otimes_{1}} \Phi_{n},
\,\,\,\, 
a(w)^+ \Phi_n = \overline{w} \widehat{\otimes} \Phi_{n}.
\]  
Note that for any element $e_1 \otimes \cdots \otimes e_n \in E^{\otimes n}$ and $w\in E^*$ 
we have well-defined right-contraction 
$w \otimes_1 \left[e_1 \otimes \cdots \otimes e_n\right] = \langle w,e_n \rangle \, e_1 \otimes \cdots \otimes e_{n-1}$
where $\langle w,e_n \rangle$ is the value of the functional $w \in E^*$ at the test element $e_n \in E$ (\emph{i.e.} dual pairing), 
which becomes
equal to the inner product $(\overline{w},e_1)_{{}_{\mathcal{H}}}$,
if $w \in E \subset E^*$. This formula of the contraction $\otimes_1$ uniquely extends over $E^* \times E^{\otimes n}$.
Its final symmetrization or alternation (for the Bose, respectively, Fermi case) defines the symmetrized/alteranted
contraction $\widehat{\otimes_{1}}$ used above \cite{obataJFA}. The structure of the Gelfand triple in the Fock space allows us to 
introduce creation-annihilation operators $a(w)^+,a(w)$ of the particles in the distributional states $w \in E^*$, which
are well-defined operators mapping continuously the Hida test space $(E)$ into its strong dual $(E)^*$ \cite{obataJFA}. 
In particular, the canonical Hida creation-annihilaion operators are defined to be equal 
to the creation-annihilation operators of the particles in the states with the spin-momentum exactly equal 
$\boldsymbol{\p}$, \emph{i.e.}  
$\partial_{\boldsymbol{\p}}^{+} {:} = a(\delta_{{}_{\boldsymbol{\p}}})^+$, 
$\partial_{\boldsymbol{\p}} {:} = a(\delta_{{}_{\boldsymbol{\p}}})$ with $\delta_{{}_{\boldsymbol{\p}}} \in E^*$
being equal to the Dirac delta functional centered at $\boldsymbol{\p}$. The reader may now understand why we need standard realization of the 
Gelfand triples, because otherwise Dirac delta would be meaningless. In fact we use the fact that each element of $E$
(a.e. equality equivalence class) has unique continuous (even smooth) representant, and, in the further development of the theory,
we use also the continuity $\boldsymbol{\p} \rightarrow \delta_{{}_{\boldsymbol{\p}}} \in E^*$.  Note that 
$\partial_{\boldsymbol{\p}}^{+}, \partial_{\boldsymbol{\p}}$ are not only formal symbolic distributional kernels 
which make sense only after smearing  out with test functions of  $\boldsymbol{\p}$ (as in the Wightman approach)
but for each particular $\boldsymbol{\p}$ they are well-defined generalized operators. 
But this not the most important difference in comparison to the Wightman definition. Having given the test
Hida space $(E)$ and its dual $(E)^*$ we can consider generalized operators of the class 
$\mathscr{L}\left(\mathscr{E},\mathscr{L}((E), (E)^*)\right) = \mathscr{L}\left((E)  \otimes \mathscr{E}, (E)^*\right)$ including the generalized integral kernel operators  
\begin{equation}\label{Xi(xi(x))}
\Xi\left((\kappa_{\mathpzc{l}\mathpzc{m}}(x) \right) =
\int \kappa_{\mathpzc{l}\mathpzc{m}}\big(\boldsymbol{\p}_1, \ldots, \boldsymbol{\p}_\mathpzc{l},
\boldsymbol{\q}_1, \ldots, \boldsymbol{\q}_\mathpzc{m};x\big) \,\,
\partial_{\boldsymbol{\p}_1}^{+} \ldots \partial_{\boldsymbol{\p}_\mathpzc{l}}^{+} 
\partial_{\boldsymbol{\q}_1} \ldots \partial_{\boldsymbol{\q}_\mathpzc{m}} 
d\boldsymbol{\p}_1 \ldots d\boldsymbol{\p}_\mathpzc{l}
d\boldsymbol{\q}_1 \ldots d\boldsymbol{\q}_\mathpzc{m}
\end{equation}        
with \emph{any} $\mathscr{E}^*$-valued distributional kernels 
$\kappa_{\mathpzc{l}\mathpzc{m}} \in \mathscr{L}\left(E^{\otimes (\mathpzc{l}+\mathpzc{m})}, \mathscr{E}^*\right)$
for any  countably Hilbert nuclear space $\mathscr{E}$ \cite{obataJFA}, of which the free field operators (\ref{FreeField=Xi(xi01)+Xi(xi10)})
are only particular cases. Each such operator  (\ref{Xi(xi(x))}) is uniquely determined by the following equality
of the dual pairings 
\[
\left\langle\left\langle \Xi(\kappa_{\mathpzc{l}\mathpzc{m}})(\Phi\otimes\phi) \right\rangle\right\rangle
= \left\langle \kappa_{\mathpzc{l}\mathpzc{m}}(\phi), \eta_{{}_{\Phi,\Psi}} \right\rangle,
\,\,\,\, \Phi,\Psi \in (E), \,\, \phi\in\mathscr{E},
\]
where $\langle \cdot, \cdot \rangle$, $\langle\langle \cdot, \cdot \rangle\rangle$, are dual pairings between $E^*$ and $E$ and,
respectively, between $(E)^*$ and $(E)$, and where the function $\eta_{{}_{\Phi,\Psi}}$ 
\[
 \eta_{{}_{\Phi,\Psi}}(\boldsymbol{\p}_1, \ldots, \boldsymbol{\p}_\mathpzc{l},
\boldsymbol{\q}_1, \ldots, \boldsymbol{\q}_\mathpzc{m}) {:}= \left\langle\left\langle\partial_{\boldsymbol{\p}_1}^{+} \cdots \partial_{\boldsymbol{\p}_\mathpzc{l}}^{+}
\partial_{\boldsymbol{\q}_1} \cdots \partial_{\boldsymbol{\q}_\mathpzc{m}}\Phi, \Psi \right\rangle\right\rangle,
\,\,\,  \Phi,\Psi \in (E)
\]
always belongs to $E^{\otimes(\mathpzc{l}+\mathpzc{m})}$ \cite{obataJFA}. 
In particular, we can construct the Wick product of the operators of the type  
(\ref{FreeField=Xi(xi01)+Xi(xi10)}), which again is a finite sum of integral kernel operators of the type (\ref{Xi(xi(x))}),
with the kernels being equal to the pointwise products of the kernels of the Wick factors (\ref{FreeField=Xi(xi01)+Xi(xi10)}). Also the
(tensor) product $W(x)W(y)$ of the Wick products of free fields is again equal to a finite sum of the integral kernel operators 
of the type  (\ref{Xi(xi(x))}) with the kernels defined by the products of the kernels of the factors $W(x)$ and $W(y)$,
including the contractions of the product kernels with respect to the corresponding spin-momenta variables, which are expressed through
absolutely convergent integrals \cite{WN}. This assertion is a rigorous equivalent of the ``Wick theorem for products'' \cite{Bogoliubov_Shirkov}. 
Thus, in particular, the Wick theorem for products can be transferred into a subclass of integral kernel operators with 
vector valued kernels. It is important that the class of finite sums of integral kernel operators with vector-valued kernels
admitting the operation of (tensor) product  includes all Wick products of free fields with coefficients equal to any translationally 
invariant tempered distributions. We should emphasize
that these results are valid for any general mixed Fock space of Bose and Fermi fields equal to the tensor product of the Fock spaces 
of the particular fields and
which can be realized as the Fock space over the total single particle space being equal to the corresponding direct sum of the particular
kinds of the free fields, with the standard operator of the total single particle space being equal to the direct sum of the standard
operators of the particular single particle spaces. This is nothing else but the general Fock lifting 
(without any symmetrizations/alternations) in which we finally symmetrize/alternate all spin-momenta variables 
corresponding to one and the same Bose/Fermi field. For any general mixed Fock space (including finite number of Bose and Fermi fields) 
the class of finite sums of integral kernel operators with vector-valued kernels
admitting the operation of (tensor) product  includes all Wick products of free fields with coefficients equal to any translationally 
invariant tempered distributions \cite{WN}.  This theorem allows to formulate the perturbative QFT with the axioms (I)-(IV) for $S$,  and with the 
free field and $S_n$ operators regarded as finite sums of integral kernel operators  (\ref{Xi(xi(x))}) with vector valued kernels
(in general with several space-time variables) \cite{WN}.

The reader can see now a general difference between the class of generalized integral kernel operators (\ref{Xi(xi(x))}) with $\mathscr{E}^*$ valued distributional 
kernels $\kappa_{\mathpzc{l}\mathpzc{m}}$ and the class of generalized 
operators in the Wightman sense. The operator valued distribution in the Wightman sense, when smeared
out with a space-time test function $\phi \in \mathscr{E}$, when expressed in the normal-order product form analogous to   
(\ref{Xi(xi(x))}), puts rather strong condition on $\kappa_{\mathpzc{l}\mathpzc{m}}$ evaluated at $\phi$, so that
$\kappa_{\mathpzc{l}\mathpzc{m}}(\phi)$ should represent a normalizable $\mathpzc{l}+\mathpzc{m}$ particle state as the spin-momenta
function, rapidly decreasing in these variables. But for the generalized integral kernel operator (\ref{Xi(xi(x))})
to be well-defined, it is sufficient that $\kappa_{\mathpzc{l}\mathpzc{m}}(\phi)$ represents (not necessary normalizable)
$\mathpzc{l}+\mathpzc{m}$ particle generalized state, in the spin-momenta variables, \emph{i.e.} it is sufficient
that $\kappa_{\mathpzc{l}\mathpzc{m}}(\phi)$ is a distribution in $E^{*\otimes (\mathpzc{l}+\mathpzc{m})}$, continously depending
on $\phi$, because, by the kernel theorem, $\kappa_{\mathpzc{l}\mathpzc{m}} \in \mathscr{L}\left(E^{\otimes (\mathpzc{l}+\mathpzc{m})}, \mathscr{E}^*\right)
\simeq \mathscr{L}\left(\mathscr{E}, E^{* \otimes (\mathpzc{l}+\mathpzc{m})}\right)$, concerning linear structure and topology;
compare thms. 3.6, 3.9 of \cite{obataJFA}.
These circumstances alone show that despite the lack of any adiabatic limit in the sense of Wightman operators, this limit can still 
exist in the sense of generalized operators with vector-valued kernels in the sense of \cite{obataJFA}.

Having given the perturbative QFT with the axioms (I)-(V) and with Hida operators \cite{WN}, let us briefly present the general results,
which have been achieved in this theory, and present some further perspectives.

Introduction of the Hida operators into the Bogoliubov, Epstein, Glaser construction of the scattering operator
converts the $n$-th order contributions $S_n(g^{\otimes \, n})$
and $W_{{}_{\textrm{int}}}^{(n)}(g_{{}_{0}}^{\otimes \, n},\phi)$ 
to the scattering operator  
and to the interacting Wick product fields $W_{{}_{j \,\, \textrm{int}}}(g_{{}_{0}},\phi)$
into the finite sums of generalized integral kernel operators $\Xi(\kappa_{\mathpzc{l}\mathpzc{m}})$:
\begin{multline}\label{Sn(g)}
S_n(g^{\otimes \, n}) = \sum\limits_{\mathpzc{l},\mathpzc{m}}
\int \kappa_{\mathpzc{l}\mathpzc{m}}\big(\boldsymbol{\p}_1, \ldots, \boldsymbol{\p}_\mathpzc{l},
\boldsymbol{\q}_1, \ldots, \boldsymbol{\q}_\mathpzc{m};  g^{\otimes \, n} \big) \,\,
\partial_{\boldsymbol{\p}_1}^{+} \ldots \partial_{\boldsymbol{\p}_\mathpzc{l}}^{+} 
\partial_{\boldsymbol{\q}_1} \ldots \partial_{\boldsymbol{\q}_\mathpzc{m}} 
d\boldsymbol{\p}_1 \ldots d\boldsymbol{\p}_\mathpzc{l}
d\boldsymbol{\q}_1 \ldots d\boldsymbol{\q}_\mathpzc{m}
\\
=
\sum\limits_{\mathpzc{l},\mathpzc{m}} \Xi\left((\kappa_{\mathpzc{l}\mathpzc{m}}(g^{\otimes \, n}) \right)
=
\sum\limits_{j_1,\ldots,j_n=0}^{k}\bigintsss \ud^4 x_1 \ldots \ud^4x_n \, S_n(j_1,x_1, \ldots, j_n, x_n) \, g_{{}_{j_1}}(x_1) \ldots g_{{}_{j_n}}(x_n),
\end{multline}
and
\begin{multline}\label{Wjint}
W_{{}_{j \,\, \textrm{int}}}^{(n)}(g_{{}_{0}}^{\otimes \, n},\phi) = \sum\limits_{\mathpzc{l},\mathpzc{m}}
\int \kappa_{\mathpzc{l}\mathpzc{m}}\big(\boldsymbol{\p}_1, \ldots, \boldsymbol{\p}_\mathpzc{l},
\boldsymbol{\q}_1, \ldots, \boldsymbol{\q}_\mathpzc{m};  g_{{}_{0}}^{\otimes \, n} \otimes \phi \big) \,\,
\partial_{\boldsymbol{\p}_1}^{+} \ldots \partial_{\boldsymbol{\p}_\mathpzc{l}}^{+} 
\partial_{\boldsymbol{\q}_1} \ldots \partial_{\boldsymbol{\q}_\mathpzc{m}} 
d\boldsymbol{\p}_1 \ldots d\boldsymbol{\p}_\mathpzc{l}
d\boldsymbol{\q}_1 \ldots d\boldsymbol{\q}_\mathpzc{m}
\\
=
\sum\limits_{\mathpzc{l},\mathpzc{m}} \Xi\left((\kappa_{\mathpzc{l}\mathpzc{m}}(g_{{}_{0}}^{\otimes \, n}\otimes \phi) \right)
=
\int \ud^4 x_1 \ldots \ud^4x_n \ud^4 x \, W_{{}_{j \,\, \textrm{int}}}^{(n)}(x_1, \ldots, x_n; x) \, g_{{}_{0}}(x_1) \ldots g_{{}_{0}}(x_n) \, \phi(x),
\end{multline}
with vector-valued distributional kernels $\kappa_{\mathpzc{l}\mathpzc{m}}$ in the sense of \cite{obataJFA}, with the values in the distributions
over the test nuclear space
\[
\left(\oplus_{0}^{k}\mathscr{E}\right)^{\otimes \, n} \ni g^{\otimes \, n} 
\,\,\,\,\,\,\,\,\,
\textrm{or, respectively,}
\,\,\,\,\,\,\,\,\,
\mathscr{E}^{\otimes \, n} \otimes (\oplus_{1}^{d}\mathscr{E}) \ni g_{{}_{0}}^{\otimes \, n} \otimes \phi
\]
with $\mathscr{E} = \mathcal{S}(\mathbb{R}^4)$.
Each of the $3$-dim Euclidean integration $d\boldsymbol{\p}_i$ with respect to the spatial momenta $\boldsymbol{\p}_i$ components
$\boldsymbol{\p}_{i1}, \boldsymbol{\p}_{i2}, \boldsymbol{\p}_{i3}$,
also includes here summation over the corresponding discrete spin components $s_i\in(1,2,\ldots)$ hidden under the symbol $\boldsymbol{\p}_i$. 

The class to which the operators $S_n$ and $W_{{}_{j \,\, \textrm{int}}}^{(n)}$ belong, expressed in terms of the Hida test space,
depend on the fact if there are massless free fields present in the interaction Lagrange density operator $\mathcal{L}$ or not.
Namely: 
\[
S_n \in
\begin{cases}
\mathscr{L}\left(\left(\oplus_{0}^{k}\mathscr{E}\right)^{\otimes \, n}, \, \mathscr{L}((E),(E))\right) \cong
\mathscr{L}\left((E), (E)\right) \otimes \mathscr{L}\left( \left(\oplus_{0}^{k}\mathscr{E}\right)^{\otimes \, n}, \mathbb{C} \right), 
& \text{if all fields in $\mathcal{L}$ are massive},\\
\mathscr{L}\left(\left(\oplus_{0}^{k}\mathscr{E}\right)^{\otimes \, n} , \, \mathscr{L}((E),(E)^*)\right)
\cong \mathscr{L}\left((E), (E)^* \right) \otimes \mathscr{L}\left( \left(\oplus_{0}^{k}\mathscr{E}\right)^{\otimes \, n}, \mathbb{C} \right), 
& \text{if massless fields are in $\mathcal{L}$}.\\
\end{cases}
\]
Because each skew-symmetric tempered distribution also is a continuous Grassmann-valued functional on the \emph{Grassmann test function space}
\cite{WN}, then causal perturbative method makes rigorous sense also 
in case, some Wick products $W_{{}_{j}}$ are odd in Fermi fields, 
with the corresponding test components $g_{{}_{j}}$ replaced with Grassmann test functions. In this case
\[
S_{n} \in
\begin{cases}
\underset{r+p=n}{\oplus}\mathscr{L}\left((E), (E)\right) \otimes \mathscr{L}\left(\left(\oplus_{0}^{k}\mathscr{E}\right)^{\otimes \, r} 
\otimes \mathscr{E}^p, \mathcal{E}^{p \, *}\right), 
& \text{if all fields in $\mathcal{L}$ are massive},\\
\underset{r+p=n}{\oplus}\mathscr{L}\left((E), (E)^*\right) \otimes \mathscr{L}\left(\left(\oplus_{0}^{k}\mathscr{E}\right)^{\otimes \, r} 
\otimes \mathscr{E}^p, \mathcal{E}^{p \, *}\right), 
& \text{if there are massless fields in $\mathcal{L}$},\\
\end{cases}
\]
with $\mathcal{E}^{p \, *}$ being the subspace of grade $p$ of the \emph{abstract Grassmann algebra} $\oplus_p\mathcal{E}^{p \, *}$ 
\emph{with inner product and involution} in the sense of \cite{Berezin}, \cite{WN}. 
$\mathcal{E}^p$ denotes the space of \emph{Grassmann-valued test functions} $g^p$ of grade $p$ 
due to \cite{Berezin}, and replacing ordinary test functions $g^{\otimes \, p}$, compare \cite{WN}.
Recall, that $\mathscr{L}(E_1,E_2)$ denotes the linear space of linear continuous operators 
$E_1\longrightarrow E_2$ endowed with the natural topology of uniform
convergence on bounded sets.

Existence of the product operation in the \emph{whole} class 
$\mathscr{L}\left((E), (E)^* \right) \otimes \mathscr{L}\left(\mathscr{E}^{\otimes \, n},\mathbb{C}\right)$
or $\mathscr{L}\big((E), (E)^*\big) \otimes \mathscr{L}(\mathscr{E}^{\otimes \, (n-p)} \otimes \mathscr{E}^p, \mathcal{E}^{p \, *})$ of operators
is quite not obvious. But the higher order contributions $S_n$ to the scattering operator, which also define the interacting fields,
are of special class, and admit the operation of product defined by the limit operation in which we replace the massless kernels
$\kappa_{0,1},\kappa_{1,0}$ of the massless fields by their massive counterparts and pass to the zero mass limit \cite{WN}, so that e.g. the axiom
(I) makes sense when we are using Hida operators.

\section{Application to QED}\label{QEDapplication}

Let us give examples of applications of this general perturbative approach, 
based on (I)-(V) with Hida operators, to the realistic QFT's, staring with spinor and scalar QED.

In spinor QED, we consider the following Krein-self adjoint interaction Lagrangian (\ref{L})
with the switching-off function $g = (g_{{}_{0}}, g_{{}_{1}}, \ldots, g_{{}_{12}}) = (g_{{}_{0}}, h_{a}, h_{b}, j_\mu)$,
$a,b \in \{1,\ldots, 4\}$, $\mu \in\{0,\ldots,3\}$, which is equal
\begin{multline*} 
\sum\limits_{j=0}^{12} g_{{}_{j}}(x) \mathcal{L}_{{}_{j}}(x)
= g_{{}_{0}}(x) \mathcal{L}_{{}_{0}}(x)  + h(x)^\sharp \boldsymbol{\psi}(x) 
+  \boldsymbol{\psi}^\sharp(x) h(x) + j(x)A(x)
\\
=  g_{{}_{0}}(x) \mathcal{L}_{{}_{0}}(x)  + \sum\limits_{a} \overline{h_{a}(x)} \big[\gamma^0 \boldsymbol{\psi}\big]^{a}(x)
 + \sum\limits_{a} h_b(x) \boldsymbol{\psi}^{\sharp \, b}(x) + \sum\limits_{\mu} j_{\mu}(x) A^{\mu}(x),
\end{multline*}
with the four component  bispinor switching-off function 
\begin{equation}\label{h=iota.phi}
h^a(x) = h^a(x)  = \iota_a(x)\phi^a(x) = \iota \cdot \phi^a(x), \,\,\,\, \phi\in \mathcal{S}(\mathbb{R}^4; \mathbb{C}^4),
\end{equation}
whose components are equal to the \emph{generators} $\iota_1(x), \ldots, \iota_4(x), \,\,\,\, x\in\mathbb{R}^4$,
\emph{of the Grassmann algebra with inner product and with involution $\overline{\,\, \cdot \,\,}$} in the sense of \cite{Berezin},
multiplied, respectively, by the Schwartz test functions $\phi^1(x), \ldots, \phi^4(x)$.

Using the notation introduced above, it is easily seen that, with the set $Z=\{x_1,x_2, \ldots, x_n\}$ of $(j=0)$-type variables,
and with $x$ being of $j$-type, the distributional kernels  of  $W_{{}_{j \,\, \textrm{int}}}^{(n)}$ are equal
\[
W_{{}_{j \,\, \textrm{int}}}^{(n)}(x_1, \ldots, x_n;x) = \textstyle{\frac{1}{i}} A_{n+1}(Z,x) 
= \textstyle{\frac{1}{i}} \textrm{adv} \, D_{n+1}(Z,x).
\]
Some of the causal distributional scalar coefficients in the Wick decomposition 
of $D_{n+1}$ have non-negative singularity order and, thus, their splitting into advanced and retarded parts, in the computation of the advanced part 
of $D_{n+1}$, is correspondingly non-unique. It can be shown that the choice in the splitting of the causal 
distributions encountered in the computation of the interacting Dirac and e.m. potential
fields $\psi_{{}_{\textrm{int}}}^{(n)}(g_{{}_{0}}^{\otimes \, n})$ 
and $A_{{}_{\textrm{int}}}^{(n)}(g_{{}_{0}}^{\otimes \, n})$, is equivalent to the choice of the splitting of the causal
distributions we encounter in the computation of the scattering matrix $S(g_{{}_{0}})$ corresponding to the Lagrangian QED interaction
$\mathcal{L} = \mathcal{L}_{{}_{0}}$ without any additional terms in the generalized Lagrangian. 
There is one \emph{natural} or  \emph{on mass shell} normalization of the splitting in QED which fixes the splitting in computation
of $S(g_{{}_{0}})$, defined in the following way. Consider the total contribution 
coming from the sum of all strongly connected graph contributions to $1/n!S(x_1, \ldots, x_n)$ which, 
when integrated with respect to all intermediate $x_3, \ldots, x_{n}$,
is of the form $-i\Pi^{\mu \nu}(x_1-x_2){:}A_\mu(x_1)A_{\nu}(x_2){:}$. It is called the 
``vacuum polarization'' contribution. Consider analogously the total contribution
coming from the sum of all strongly connected graph contributions to $1/n!S(x_1, \ldots, x_n)$ which, 
when integrated with respect to all intermediate $x_3, \ldots, x_{n}$, are of the form 
$-i\Sigma^{ab}(x_1-x_2){:}\psi^{\sharp}_{a}(x_1)\boldsymbol{\psi}_{b}(x_2){:}$. It is called the 
``self energy term''. Let $\widetilde{\Pi^{\mu \nu}}$, $\widetilde{\Sigma}$ be the Fourier transforms of $\Pi^{\mu \nu}, \Sigma$.  
The \emph{natural} or  \emph{on mass shell} normalization is defined by the following conditions valid in each order:
\begin{multline}\label{OnShell}
\widetilde{\Pi^{\mu \nu}}(0) = 0, \,\,\,
\textstyle{\frac{1}{p^2}}\widetilde{\Pi^{\mu \nu}}(p)\Big{|}_{{}_{p^2=0}} = 0,
\\
(\slashed{p}+m)\widetilde{\Sigma}(p)\Big{|}_{{}_{p^2=m^2}} = \widetilde{\Sigma}(p)(\slashed{p}+m)\Big{|}_{{}_{p^2=m^2}} = 0, \,\,\,
\,\,\,
\textstyle{\frac{1}{m-\slashed{p}}}\widetilde{\Sigma}(p)\Big{|}_{{}_{p^2=m^2}} 
= \widetilde{\Sigma}(p)\textstyle{\frac{1}{m-\slashed{p}}}\Big{|}_{{}_{p^2=m^2}} = 0, 
\end{multline}
where $\left(m-\slashed{p}\right)^{-1} = \tfrac{m+\slashed{p}}{m^2-p^2-i\epsilon} = \widetilde{S^c}(p)$ and $m$ being the electron mass.

Let $g_{{}_{0 \, \epsilon}}(x) {:}= g_{{}_{0}}(\epsilon x)$, for a fixed $g_{{}_{0}} \in \mathscr{E}$, $g_{{}_{0}}(0) = \alpha_{{}_{0}}$. 
In particlular the set of functions $\{g_{{}_{0 \, \epsilon}}, \epsilon \in \mathbb{R}_+\}$ is bounded in $\mathscr{E}$ and 
$g_{{}_{0 \, \epsilon}} \rightarrow const = \alpha_{{}_{0}}$ almost uniformly when $\epsilon \rightarrow  0^+$. 
In each case we require existence of the numerical limit 
\begin{equation}\label{WeakAL}
\underset{\epsilon \rightarrow 0^+}{\textrm{lim}}{\langle\langle S_n(g_{{}_{0 \, \epsilon}}^{\otimes n}) \Phi_0, \Phi_0\rangle\rangle},
\end{equation} 
where $\Phi_0$ is the vacuum, and its independence of the test unction $g_{{}_{0}}$,
which is equivalent to existence of (\ref{WeakAL}) and its equality to $0$, for all $n>1$, so that 
we have the normalization 
\begin{equation}\label{NS}
\underset{\epsilon \rightarrow 0^+}{\textrm{lim}}{\langle\langle S(g_{{}_{0 \, \epsilon}}) \Phi_0, \Phi_0\rangle\rangle} =1.
\end{equation}
This normalization, together with (\ref{OnShell}) and gauge invariance, determine the splitting uniquely.

We have the follwing theorems \cite{IF}: 

\begin{twr}
If the electron mass $m\neq 0$, then there exists unique normalization of the splitting, 
called \emph{natural} or \emph{on mass shell normalization}, such that 
\[
\underset{\epsilon \rightarrow 0^+}{\textrm{lim}}\psi_{{}_{\textrm{int}}}^{(n)}(g_{{}_{0 \, \epsilon}}^{\otimes \, n})
= \psi_{{}_{\textrm{int}}}^{(n)},
\,\,\, 
\underset{\epsilon \rightarrow 0^+}{\textrm{lim}}A_{{}_{\textrm{int}}}^{(n)}(g_{{}_{0 \, \epsilon}}^{\otimes \, n})
= A_{{}_{\textrm{int}}}^{(n)}
\]
exist as finite sums of generalized integral kernel operators with vector valued kernels in the natural uniform topology on bounded sets in 
\begin{equation}\label{ClassIntFields}
\mathscr{L}\big(\oplus_{1}^{4}\mathscr{E}, \, \mathscr{L}((E),(E)^*)\big), \,\,\,\,\phi \in \oplus_{1}^{4}\mathscr{E}.
\end{equation}
\label{electronmass=/=0}
\end{twr}

\begin{twr}
If the electron mass $m=0$, then for each choice of the normalization of the splitting, the limits
\[
\underset{\epsilon \rightarrow 0^+}{\textrm{lim}}\psi_{{}_{\textrm{int}}}^{(n)}(g_{{}_{0 \, \epsilon}}^{\otimes \, n})
\,\,\, 
\underset{\epsilon \rightarrow 0^+}{\textrm{lim}}A_{{}_{\textrm{int}}}^{(n)}(g_{{}_{0 \, \epsilon}}^{\otimes \, n})
\]
do not exist in the natural uniform topology on bounded sets in 
\begin{equation}\label{ClassIntFields}
\mathscr{L}\big(\oplus_{1}^{4}\mathscr{E}, \, \mathscr{L}((E),(E)^*)\big), \,\,\,\,\phi \in \oplus_{1}^{4}\mathscr{E}.
\end{equation}
\label{electronmass=0}
\end{twr}

It should be stressed that if $m\neq 0$, then each normalization of the splitting, which is not \emph{natural}, gives 
higher order contributions to interacting fields which do not allow existence of the adiabatic limit. In fact the statement of the first 
theorem can be strengthened \cite{IF}: each contribution to interacting field coming from a connected graph not only converges 
in the above sense, but it is a well-defined generalized operator just with $g_{{}_{0}}$ put equal $const$.

From the results of \cite{WN} it also follows
\begin{twr}
There exists the (tensor) product operation for the adiabatic limits of  higher-order contributions to  interacting fields, whenever the limits 
exist, as finite sums of integral kernel operators with vector-valued kernels. 
\label{product}
\end{twr}
Indeed, the product operation of higher-order contributions to interacting fields in the adiabatic limit
is given through the natural formula
\[
\left(\underset{\epsilon \rightarrow 0}{\textrm{lim}} A_{{}_{\textrm{int}}}^{(n)}(g_{{}_{0 \, \epsilon}}^{\otimes \, n};x)\right)
\left( \underset{\epsilon \rightarrow 0}{\textrm{lim}} A_{{}_{\textrm{int}}}^{(m)}(g_{{}_{0 \, \epsilon}}^{\otimes \, m};y)\right),
\,\,\, \left(\underset{\epsilon \rightarrow 0}{\textrm{lim}}\psi_{{}_{\textrm{int}}}^{(n)}(g_{{}_{0 \, \epsilon}}^{\otimes \, n};x)\right)
\left( \underset{\epsilon \rightarrow 0}{\textrm{lim}}\psi_{{}_{\textrm{int}}}^{(m)}(g_{{}_{0 \, \epsilon}}^{\otimes \, m};y)\right),
\,\,\,
\ldots
\] 
where the kernels of the products are equal to the tensor products of the kernels of the limit operators \cite{WN}.

The adiabatic limit for the higher order contributions to the scattering operator $S(g_{{}_{0}})$ corresponding to the Lagrangian 
$\mathcal{L}_{{}_{0}}$ without any additional terms,  
also exists, but the limit $\underset{\epsilon \rightarrow 0^+}{\textrm{lim}}{S_n(g_{{}_{0 \, \epsilon}}^{\otimes n})}$, in the topology of uniform convergence on 
bounded sets in the space $\mathscr{L}\left(\mathscr{E}^n, \mathscr{L}((E), (E)^*)\right)$, can possibly be non-unique. This eventual non-uniqueness
is still under investigation, but it can be fixed by concrete choice of sufficiently ``adiabatic'' character of the convergence 
$g_{{}_{0}} \rightarrow const$. In fact: the kernels $\kappa_{\mathpzc{l}, \mathpzc{m}}$ of $S_n$ belong to 
$\mathscr{L}\left(\mathscr{E}, E^{*\otimes(\mathpzc{l}+\mathpzc{m})}\right)$, 
and transform the bounded set $\{g_{{}_{0 \, \epsilon}}, \epsilon \leq \epsilon_0 >0 \} \subset \mathscr{E}$ into a bounded set 
in $E^{*\otimes(\mathpzc{l}+\mathpzc{m})}$.
Because each bounded set in $E^{*\otimes(\mathpzc{l}+\mathpzc{m})}$ is relatively compact \cite{GelfandII}, there exists a subsequence 
$\kappa_{\mathpzc{l}, \mathpzc{m}}(g_{{}_{0 \, 1/k}}^{\otimes n})$ convergent to an element 
$\kappa_{\mathpzc{l}, \mathpzc{m}}(\alpha_{{}_{0}}) \in E^{*\otimes(\mathpzc{l}+\mathpzc{m})}$ which, by thm. 3.9 of \cite{obataJFA}, 
represents a well-defined generalized integral kernel operator in $\mathscr{L}((E), (E)^*)$ -- the limit contribution to the scattering operator 
$S(g_{{}_{0}} = \alpha_{{}_{0}})$. The differences between possible limits (obtained by different choices of the subsequences $g_{{}_{0 \, 1/k}}$,
$k=1,2, \ldots$) are irrelevant for the effective cross-sections for the many-particle generalized plane wave $in$ and $out$ states,
\cite{Bogoliubov_Shirkov}, \S 24.5. 

The \emph{natural} normalization (\ref{OnShell}) is nothing else but the ordinary \emph{on mass shell} normalization 
in spinor QED \cite{Bogoliubov_Shirkov}, \S 34.4, (51),(52), (45), (46), which -- concerning $\widetilde{\Sigma}$ -- 
is written frequently with the help of the ``formal derivative $d\slash d\slashed{p}$'': 
$\widetilde{\Sigma}(p)= d\widetilde{\Sigma}(p)\slash  d\slashed{p} = 0$ at $\slashed{p}=m$. It means that the complete Green functions
of the photon and the electron and positron all have, respectively, the poles at the same point as the free Green functions, and
without any radiative contributions to the external lines: with the creation-annihilation operators of the free fields interpreted as 
creation-annihilation operators of the real particles, with the charge in the (renormalized) interaction Lagrangian equal
to the real charge of the electron, and with masses of the electron, positron and photon in the free part of the (renormalized)  
Lagrangian equal to the  masses of the real particles.   

We have analogous results for the scalar QED, in which the adiabatic limit for the interacting fields exists only if the mass of charged
particles-antiparticles is non-zero, and only if the normalization is \emph{on mass shell}.

The essence of the argument in the proof \cite{IF} of the above theorems comes from the fact that the kernels of the higher order
contributions to interacting fields $\kappa_{\mathpzc{l}\mathpzc{m}}(\phi)$, evaluated at a space-time test function $\phi$,
and with $g_{{}_{0}} = const.$, exist as distributions on the $\mathpzc{l}+\mathpzc{m}$-particle test function space
$E^{\otimes (\mathpzc{l}+\mathpzc{m})}$, and continuously depend on $\phi$, provided only the normalization is
``on mass shell''. This, by thms. 3.6, 3.9 of \cite{obataJFA}, is sufficient for the existence of the integral 
kernel operator $\Xi(\kappa_{\mathpzc{l}\mathpzc{m}})$ corresponding to the kernel $\kappa_{\mathpzc{l}\mathpzc{m}}$. 
In the approach based on Wightman operator distributions existence of the adiabatic limit is impossible, because these
kernels  $\kappa_{\mathpzc{l}\mathpzc{m}}(\phi)$ are in general non-normalizable $\mathpzc{l}+\mathpzc{m}$-particle states and in general are not
rapidly decreasing Schwartz functions of the momenta.

Let us shortly explain the role of the ``on mass shell'' normalization
for the existence of the limit, understood as integral kernel operators. In the higher order contributions
to interacting $\psi_{{}_{\textrm{int}}}^{(n)}(x_1, \ldots, x_n;x)$ and $A_{{}_{\textrm{int}}}^{(n)}(x_1, \ldots, x_n;x)$, 
we have the Green functions $\Pi_{{}_{\textrm{av}, \, \textrm{ret}}}^{\mu \nu}$,
$\Sigma_{{}_{\textrm{av}, \, \textrm{ret}}}^{ab}$, immediately related to $\Pi^{\mu \nu}$,
$\Sigma^{ab}$, and equal (in the $n$-th order), respectively, to the retarded/advanced parts of the contributions to the mentioned above causal distributions 
$D_n(x_1, \ldots,x_n)$ and which are proportional, respectively, to ${:}A_\mu(x_1)A_{\nu}(x_2){:}$ or ${:}\psi^{\sharp}_{a}(x_1)\psi_{b}(x_2){:}$
(with all other coordinates integrated out). The higher order contributions to interacting fields 
can be represented by (amputed at $x$) graphs, analogously to the ordinary Feynman graphs for the ordinary scattering operator (without additional terms in the Lagrangian), and $\Pi_{{}_{\textrm{av}, \, \textrm{ret}}}^{\mu \nu}$,
$\Sigma_{{}_{\textrm{av}, \, \textrm{ret}}}^{ab}$, play the role in the higher order contributions to interacting fields
analogous as the ordinary propagators $\Pi, \Sigma$ do in the higher order contributions to the ordinary $S$-matrix. Among the higher order contributions
to interacting $A^\mu$ in the adiabatic limit $g_{{}_{0}} \rightarrow const.$, we have convolutions $D_{{}_{0}}^{{}^{\textrm{av}, \, \textrm{ret}}} \ast \Pi_{{}_{\textrm{av}, \, \textrm{ret}}}^{\mu \nu} \ast A_{\nu}$,
$D_{{}_{0}}^{{}^{\textrm{av}, \, \textrm{ret}}} \ast \Pi_{{}_{\textrm{av}, \, \textrm{ret}}}^{\mu \nu} 
\ast D_{{}_{0}}^{{}^{\textrm{av}, \, \textrm{ret}}} {:}\psi^{\sharp}\gamma_\nu \psi{:}$. Analogously, among the higher order contributions
to interacting $\psi$ we have convolutions $S^{{}^{\textrm{ret}, \, \textrm{av}}} \ast \Sigma_{{}_{\textrm{av}, \, \textrm{ret}}}^{\mu \nu} \ast \psi$,
$S^{{}^{\textrm{ret}, \, \textrm{av}}} \ast \Sigma_{{}_{\textrm{av}, \, \textrm{ret}}}^{\mu \nu} \ast S^{{}^{\textrm{ret}, \, \textrm{av}}} \ast 
{:}\gamma^\nu\psi A_\nu{:}$. Because the convolutions turn into products under the Fourier transform, we would get into trouble
with these terms because the Fourier transforms of the kernels of the free fields $A, \psi$ are concentrated at the the corresponding
mass shell: $p\cdot p =0$ or $p\cdot p = m^2$, similarly as the Fourier transforms of the corresponding
ret and av parts of the commutation functions $D_{{}_{0}}^{{}^{\textrm{av}, \, \textrm{ret}}}, S^{{}^{\textrm{ret}, \, \textrm{av}}}$,
so that the above-mentioned convolutions would be ill-defined, unless the normalization is ``on mass shell'' in which 
$\Pi_{{}_{\textrm{av}, \, \textrm{ret}}}^{\mu \nu}$,
$\Sigma_{{}_{\textrm{av}, \, \textrm{ret}}}^{ab}$, respect the same conditions (\ref{OnShell}), which in fact are
equivalent to (\ref{OnShell}) for $\Pi^{\mu\nu}, \Sigma$. Indeed, denoting the kernels of the free field (\ref{FreeField=Xi(xi01)+Xi(xi10)}) 
for the Dirac field $\mathbb{A} = \psi$, evaluated at the single particle
test function $\xi \in E$, by $\kappa_{0,1}(\xi), \kappa_{1,0}(\xi)$, the value of the
contribution $S^{{}^{\textrm{ret}}} \ast \Sigma_{{}_{\textrm{ret}}}^{\mu \nu} \ast \psi$ at the space-time test function $\phi$,
is equal to the sum of two contributions, the negative frequency one:
\[
\big\langle S^{{}^{\textrm{ret}}} \ast \Sigma_{{}_{\textrm{ret}}}^{\mu \nu} \ast  \kappa_{0,1}(\xi), \phi \big\rangle
= \underset{\epsilon \rightarrow 0^+}{\textrm{lim}} \sum\limits_{s=1}^{2}\int d\boldsymbol{\p} \xi_s(\boldsymbol{\p})
{\textstyle\frac{(m+\slashed{p})\widetilde{\Sigma_{\textrm{ret}}}(p_0(\boldsymbol{\p}), \boldsymbol{\p})u_s(\boldsymbol{\p})}{-i\epsilon p_0(\boldsymbol{\p})}} 
\widetilde{\phi}(p_0(\boldsymbol{\p}), \boldsymbol{\p})
\]
and analogously the positive frequency one, with the positive frequency solutions $u_s$ replaced with the negative frequency solutions
$v_s$. This limit indeed exists only if $\Sigma_{{}_{\textrm{ret}}}$ respects the ``on shell condition'' (\ref{OnShell}), 
and degenerates to zero. Analogously we have for the contribution 
$D_{{}_{0}}^{{}^{\textrm{av}}} \ast \Pi_{{}_{\textrm{av}}}^{\mu \nu} \ast A_{\nu}$, which is meaningful only if $\Pi_{{}_{\textrm{av}}}$
respects the ``on mass shell'' condition (\ref{OnShell}).

The difference between the QED with massive and massless charged fields comes from the fact that in the massless case 
the Fourier transforms of the products of the pairing functions
as well as their causal combinations in the contributions to the operators $D_n$ are not analytic at the ``zero mass shell'' $p \cdot p = 0$, but have
singularity there. The same is true for their ret and av parts, which contribute to the scattering operator and to the interacting fields. In particular
the normalization point in the massless case cannot be chosen at zero, and ``on mass shell normalization'' becomes impossible, so that some of the
contributions  to interacting fields (e.g. mentioned above)  are not well-defined in the adiabatic limit, even as integral kernel operators 
in the white noise sense, and for no choice of the normalization.

We can summarize the results, and restate them in still another form. The perturbative QFT, with the Hida operators as the canonical
creation-annihilation operators, can be subsumed by the Bogoliubov causality axioms (I)-(V), with the freedom in the normalization
(when computing retarded and advanced components of the causal distributions $D_n$), eliminated by the axiom (VI) presupposing existence
of the limit (\ref{WeakAL}), the normalization (\ref{NS}) and existence of the adiabatic limit for higher 
order contributions to interacting fields, understood as finite sums of integral kernel operators with vector-valued kernels in the sense \cite{obataJFA}
and the axiom (VII) of gauge invariance. From perturbative QED, understood in this sense, it follows that charged particles 
are necessary massive (equivalent formulation of the above stated theorems). The ``on mass shell'' normalization is a consequence 
of the said axioms (I)-(VII).

\section{Further perspectives}\label{Perspectives}

Presented formulation of \emph{perturbative} QFT, based on Hida operators
is by no means confined to QED, but can be applied to any QFT \cite{WN}. 
But we should emphasize that presented QFT, in which the freedom in normalization of the splitting into retarded and advanced 
parts is eliminated, is \emph{perturbative}. This perturbative QFT should be supplemented and compared with non-perturbative 
methods in order to gain a better insight into the nature of QFT problems, which is indeed practiced. This is of particular importance
in the case of QFT in which the $\beta$-function remains small and negative in the vicinity of zero, \emph{i.e.} in QFT with asymptotic freedom 
(Yang-Mills fields coupled minimally to charged fields, including the case with spontaneous symmetry breaking). For such a QFT, we have 
a possibility of computing the UV asymptotics of the Green functions, based on the assumption of invariance under the renormalization group action. 
At first sight it seems that the perturbative QFT, we present here, which eliminates any freedom in renormalization, is in conflict
with the methods based on the renormalization group, in which the freedom in the choice of normalization plays a fundamental role.   
But this is not at all the case, because in the application of normalization freedom, we have to stay within the UV-asymptotics of Green functions (in QFT with asymptotic freedom) in order to stay within the range of applicability of 
the perturbative theory, where we obtain (deeply Euclidean) UV-asymptotics of Green functions. 
These results, obtained from the renormalization group invariance, 
are local in their character, concern the UV-asymptotics of Green functions,
or rather, their local behavior. In QFT's with asymptotic freedom, the IR-asymptotics of Green functions, or global behavior of Green functions
lies beyond the perturbative and renormalization group methods \cite{SlavnovFaddeev}, Chap. V.2.  
We cannot expect the adiabatic limit to exist in each particular order in the domain, which is beyond the 
range of perturbative theory, and, which heavily depends on the \emph{nonlocal} behavior (adiabatic limit) of the Green functions. 
In fact, in the ordinary massless Yang-Mills theory, our results are in complete agreement with what we already know. 
The perturbative QFT, understood in the sense presented here, 
is inapplicable to massless Yang-Mills theory, concerning the global aspects, including the adiabatic limit.
Although the argument we are using here is different, neither using the behavior of the $\beta$-function, and nor the value of the coupling constant: 
the adiabatic limit for interacting fields does not exist in massless Yang-Mills theory as the immediate consequence of the well-known
fact that the Fourier transforms of the products of the pairing functions, of their causal combinations, and of their ret and av parts are singular
at the cone $p\cdot p=0$, and the ``on mass shell'' normalization is impossible in this theory. But we can join the perturbative QFT
presented here with the non-perturbative results. The last suggests that in the range of scattering phenomena with large momentum transfer,
QFT's with asymptotic freedom can be treated perturbativly. The only natural way to safe all the axioms (I)-(VII) for perturbative 
QFT and keep the Lagrangians of the QFT's with non-abelian gauge and asymptotic freedom, 
is to use the spontaneous symmetry-breaking mechanism. Thus, the further perspective
we have in mind is the application of the presented perturbative QFT to the (massive) Yang-Mills fields coupled minimally to 
charged fields with spontaneous symmetry breaking. We already know that in such theories' fulfillment of the above axioms (I)-(VII)
with Hida operators is possible, with nontrivial mass relations coming from (I)-(VII). Therefore, it seems that asymptotic freedom, together
with the perturbative QFT, presented here, speaks for the symmetry breaking mechanism.
However, the problem requires further investigation, as we have various possibilities for the symmetry breaking,
and there are various possible realizations of the massive four-vector fields (among them using not only fermion massless ghosts, 
but also massive boson ghosts \cite{Grigore}) and all these approaches also require comparison.

\section{Comparison to other approaches}\label{Alternatives}

We should emphasize, that the adiabatic limit axiom (VI) presupposes the existence of the adiabatic limit for interacting fields 
in each order as a finite sum of generalized integral kernel operators in the sense of the natural topology
of generalized operators. Its fulfillment in QED we have proved using the Hida operators and white noise calculus (\cite{obataJFA}, \cite{WN},
\cite{IF}) for integral kernel operators -- tools which, up to the author's knowledge, have not been used before in QFT. 
In the literature, rather generalized operators in Wightman sense are used. Within this approach, the adiabatic limit axiom in the strong
sense does not exist in QED. Nonetheless, in theory based on Wightman generalized operators, a weak form of the adiabatic limit axiom 
is preserved in QED, in which the adiabatic limit exists for Green functions \cite{BlaSen}, but not for higher-order contributions 
to the interacting fields themselves. With this weak form of the adiabatic limit axiom (based on generalized operators in the Wightman sense),
the conclusions coming from the axioms are, of course, also different (weaken),
e.g. we no longer can prove that the charged particles are massive (with perturbative QFT applied to QED). 
In the case of massless Yang-Mills theory (with unbroken symmetry), the situation is similar, and we have at our disposal an alternative approach
with generalized operators in Wightman sense and a weak adiabatic limit. 
In this case, the adiabatic limit is more subtle, but physicists have learned how to deal with infrared divergences in computations
of cross-sections. This suggests that the adiabatic limit for Green functions exists also
in this case, and indeed, this existence was proved in \cite{PDuch}. Of course, within this alternative approach with
generalized operators understood in Wightman sense and weak adiabatic limit, the conclusions suggested in Section \ref{Perspectives}, 
no longer hold. Here only experiment can judge, which concrete mathematical realization of the Bogliubov axioms is correct (better),
and, as we know, this problem is experimentally open (e.g. we are waiting for further experimental mass limitations on the gluon masses). 
Since, in fact, all known electrically charged particles are massive, we can hope that the approach presented here goes in the right direction.
We also have another argument: in the case of QED, with axioms (I)-(VII) as above, we can obtain infrared asymptotics that agrees 
with the quantum theory of infrared fields developed in \cite{Staruszkiewicz}.

Now let us give a slightly more detailed comparison of our
approach with the approach based on Wightman generalized operators and the so-called weak adiabatic limit condition in perturbative 
QFT with the interaction Lagrangian $\sum\limits_{j}g_{{}_{j}}\mathcal{L}_{{}_{j}}$ with each $\mathcal{L}_{{}_{j}}$ being a Wick polynomial
in free fields, possibly containing massless fields. Let $\mathbb{A}_{{}_{1 \,\, \textrm{int}}}, \ldots, \mathbb{A}_{{}_{n \,\, \textrm{int}}}$ 
be the interacting fields -- formal power series in $g$ --  corresponding to the 
Wick products $\mathbb{A}_{{}_{1}}, \ldots, \mathbb{A}_{{}_{n}}$ of free fields. We say that the weak adiabatic limit \cite{Epstein-Glaser}, \cite{BlaSen} is fulfilled
if for the following vacuum expectation values of the products and time ordered products of interacting fields (Wightman and Green functions) 
\[
\left\langle  \Phi_0| \mathbb{A}_{{}_{1 \,\, \textrm{int}}}(g_{{}_{\epsilon}};x_1), \ldots, \mathbb{A}_{{}_{n \,\, \textrm{int}}}(g_{{}_{\epsilon}}; x_n) \Phi_0 \right\rangle
\,\,\,
\textrm{and}
\,\,\,
\left\langle  \Phi_0| T\big(\mathbb{A}_{{}_{1 \,\, \textrm{int}}}(g_{{}_{\epsilon}}; x_1), \ldots, \mathbb{A}_{{}_{n \,\, \textrm{int}}}(g_{{}_{\epsilon}};x_n)\big) \Phi_0 \right\rangle
\]
the limit $\epsilon\rightarrow 0$ exist in each order separately, where $g_{{}_{\epsilon}}(x) = \big(g_{{}_{1}}(\epsilon x), \ldots, g_{{}_{n_0}}(\epsilon x)\big)$ and 
$g_{{}_{j}} \in \mathscr{E}$, and where $g_{{}_{j}}(0) = \alpha_j$ are the coupling constants. Recall that the formal power series for the 
time ordered product of interacting fields $\mathbb{A}_{{}_{1 \,\, \textrm{int}}}, \ldots, \mathbb{A}_{{}_{n \,\, \textrm{int}}}$, 
is determined by the formal functional derivative \cite{Bogoliubov_Shirkov},\cite{Epstein-Glaser}: 
\[
T\big(\mathbb{A}_{{}_{1 \,\, \textrm{int}}}(g; x_1), \ldots, \mathbb{A}_{{}_{n \,\, \textrm{int}}}(g;x_n)\big) = 
i^n \, \left[
S(g)^{-1}
{\textstyle\frac{\delta}{\delta h_{{}_{1}}(x_1)}} \ldots {\textstyle\frac{\delta}{\delta h_{{}_{n}}(x_n)}} S(g+h)
\right]\Bigg|_{{}_{h=0}}
\]
of the formal power series for the scattering operator $S(g+h)$ corresponding to the Lagrangian 
$\mathcal{L} = \sum\limits_{j}g_{{}_{j}}\mathcal{L}_{{}_{j}}+ \sum\limits_{j} h_{{}_{k}}\mathbb{A}_{{}_{k}}$ with the ``switsching of intensity of interaction'' 
function $(g,h) = (g_{{}_{1}}, \ldots, g_{{}_{n_0}}, h_{{}_{1}}, \ldots, h_{{}_{n}})$. Thus the $m$-th order contributions to the time ordered
(advanced/retarded) interacting fields are equal to the (advanced/retarded) products \cite{Epstein-Glaser} associated to the corresponding 
scattering operator. We say that the strong adiabatic limit condition is fulfilled if 
\[
\underset{\epsilon \rightarrow 0}{\textrm{lim}} S_n(g_{{}_{\epsilon}})\Phi
\]   
exists on a dense linear subspace of states $\Phi$,
which need to satisfy certain additional invariance conditions in order, e.g., that the above limit can serve as the $n$-th order contribution 
to the scattering matrix $S$ in the adiabatic limit and in order to preserve the operator valued distribution property in the Wightman sense of the interacting
fields in the adiabatic limit \cite{Epstein-Glaser}, \cite{epstein-glaser-al}. Existence of the strong adiabitic limit was proved in
\cite{epstein-glaser-al} for the massive scalar field, and this proof can be extended on QFT containing only massive fields in the 
(renormalizable) interaction Lagrangian. The existence of the weak adiabatic limit for the purely massive theory was noted in \cite{Epstein-Glaser}, 
and is based on the fact that, in this purely masive case, the Fourier transforms of the kernels of the vacuum expectation values of the 
advanced and retarded products are zero in a neighborhood of zero 
(and which is related to the spectral condition with a mass gap). 
For perturbative QFT with infinite range of interaction and massless fields in the interaction Lagrangian,
like QED, or massless Yang-Mills theory, the strong adiabatic limit does not exist. Existence of the weak adiabatic limit was proven in \cite{BlaSen}
for QED's with massive charged fields and for the massless $\varphi^4$-theory. The method \cite{BlaSen} was subsequently extended
in \cite{PDuch} on theories for which either the canonical dimension of each $\mathcal{L}_j$ is equal to $4$ or    
the canonical dimension of each $\mathcal{L}_j$ is equal to $3$, with each monomial in each $\mathcal{L}_j$ containing at least one massive field. Thus,
the weak adiabatic limit exists also for QED's with massless charges and for the massless Yang-Mills theory (with unbroken symmetry). 
Note that, similarly to our approach based on Hida operators, the existence of the weak adiabatic limit imposes non-trivial 
conditions on the choice of normalization in the splitting of causal distributions, but much weaker ones.
However, the existence of the adiabatic limit for interacting fields in the sense of Wighman operator distributions in QED (with massive or massless charges)
or, in massless Yang-Mills theory, is impossible. Perhaps one could suppose that there is a specially distinguished linear subspace $\mathfrak{L}$ 
of normalized states on which the existence of an adiabatic limit for the interacting fields could somehow be preserved in QFT's admitting only weak
adiabatic limit. Possibly, one could have been inclined to consider a linear subspace
of states $\Phi,\Psi$ in the Fock space for which the limit for the averages in these states of the advanced or retarded products or for
\[
\underset{\epsilon \rightarrow 0}{\textrm{lim}} \left\langle \Psi|S_n(g_{{}_{\epsilon}})\Phi\right\rangle
\]
exists, to serve as $\mathfrak{L}$. But in fact no subset $\mathfrak{L}$ of this type can save the existence of this limit 
as long as we remain exclusively within the generalized operator distributions in the Wightman sense. This is because in case of QFT
with infinite range of interaction, like QED or massless Yang-Mills theory, among the higher-order contributions to the advanced products,
or to interacting fields, there are contributions which act, in the adiabatic limit, as the creation and annihilation operators of finite
number of nonnormalizable states. No subspace $\mathfrak{L}$ of normalizable states is invariant under the action of such an operator. But 
the very construction of the Wightman operator distribution requires such invariance of the domain $\mathfrak{L}$ for the construction of this distribution
\cite{wig}, \cite{Woronowicz}, compare also Sections \ref{Hida}, \ref{QEDapplication} where we have already signalized this problem. 
Otherwise, if the higher-order contribution to the interacting field is represented in the normal-order form as a finite 
sum of terms of the form (\ref{Xi(xi(x))}), then
the kernels $\kappa_{\mathpzc{l}, \mathpzc{m}}(\phi)$ in it (smeared out with space-time test functions $\phi$) 
represent in the adiabatic limit, in general, $\mathpzc{l}+\mathpzc{m}$-particle nonnormalized, generalized states. Thus, 
the higher-order contribution to interacting fields, in general, cannot be regarded as generalized operator 
in Wighman's sense, which shows that integral kernel operators with Hida operators are unavoidable here, 
because they allow $\kappa_{\mathpzc{l}, \mathpzc{m}}(\phi)$ to be a distribution continuously depending on $\phi$.     
Summing up, interacting fields and scattering operator in the adiabatic limit cannot be saved within the causal perturbative QFT approach based 
on Wightman's operator distributions and the weak adiabatic limit in QFT with massless fields, such as QED or massless Yang-Mills theory. 
An attempt is made to resolve this difficult situation by 
combining perturbative QFT with non-perturbative methods. However, all these attempts remain within the realm of hypotheses, 
trying to somewhat link this problem with the not quite clear idea of the so-called ``physical'' or ``real charged particle 
surrounded by a cloud of soft infrared photons'' (there are several approaches in this direction, 
with \cite{FaddeevKulish} among them, proposing a relatively ``small'' modification of the perturbative $S$ operator).
This seems  not entirely convincing. Let us note that, firstly, in practice, we are dealing with 
non-normalizable generalized states, as the \emph{in} and \emph{out} states in the scattering process. 
Secondly, whenever the perturbation method is also physically justified, as, e.g., in QED, the effective cross-sections, 
calculated for multi-particle generalized plane-wave states, possess adiabatic limits \cite{Bogoliubov_Shirkov},\cite{Scharf}, 
and are consistent with experiment, even though all calculations are made within the perturbation method. Therefore, it seems 
that the perturbative method is physically justified in such cases without the need for resorting to non-perturbative methods.

We therefore propose a simpler solution and remain totally within the perturbative QFT, due to Bogoliubov, Epstein and Glaser, 
except that we accept the mathematical realization of the creation-annihilation operators of the free fields as the Hida operators. 
No other modifications are introduced. This allows a natural treatment of the generalized nonnormalized states, like the plane wave states 
and the infrared states -- elements of the space dual to the Hida space $(E)$. We should emphasize here that this realization of the free fields 
introduces absolutely  no new \emph{ad hoc} structures into the theory. We emphasize this because, at first glance, it might seem that 
for example, the operator $A$ in the single-particle Hilbert space $\mathcal{H}$ (the Fock lift of which is used to construct the Hida space
$(E)$) bears the marks of a certain arbitrariness unrelated to the free field or free fields of the theory. But this is by no means the case, because,
$A$ is restricted by the single particle Gelfand triple $E\subset \mathcal{H}\subset E^*$, with
$E=\mathcal{S}(\mathbb{R}^3)$ or possibly $E=\mathcal{S}^0(\mathbb{R}^3)$ in white-noise case for massless fields.
Note that $E\subset \mathcal{H}$ is used also in the approach based on Wightman's operator distributions. It is true that
we have some arbitrariness in the choice of $A$ in the abstract Gelfand realization of this triple based on $A$, but the choice 
of any possible operator realization of this triple is completely irrelevant, even explicit form of $A$ is not relevant, 
with the only essential ingredient -- the asymptotic behavior of the spectrum of $A$, which assures that $A$ provides 
operator realization of the single-particle Gelfand triple. 
Each choice of $A$, admissible by the above requirement, gives the same Hida space $(E)$, uniquely determined by the free field(s). 
We should also emphasize, that the existence of the standard realization of the spaces $E,\mathcal{H}, E^*$, is by no means an \emph{ad hoc} assumption, 
but it is the canonical requirement in the construction of free fields. We should also warn the reader against certain subtleties regarding free fields, 
especially because their construction is often unjustly neglected and therefore poorly understood. For example, even if we put restriction
on the annihilation-creation operators to be realized over the Fock space, there are still various realizations of the massless four-vector field
in the Gupta-Bleuler gauge, having the same pairing functions, Krein-isometrically equivalent, 
but with substantially different behavior in the IR limit, e.g., with restriction of the Krein-isometric representation to the subgroup
$SL(2,\mathbb{C})$ being decomposable in the first realization and non-decomposable in the other. This non-uniqueness in the realization
of the free fields we have, irrespective of whether we regard the free fields as Wightman's operator distributions or as generalized integral kernel 
operators with vector-valued kernels (using Hida operators). In the approach based on Wighman's operator distributions and weak adiabatic limit
the difference between these various realizations of the free e.m. potential remains invisible, 
because they have the same pairing functions. But we should emphasize that using Hida operators, and regarding free fields as generalized
integral kernel operators with vector-valued kernels, we gain for free a new structure that is very important  -- the Hida's space $(E)$, composing
the Gelfand triple $(E) \subset \Gamma(\mathcal{H}) \subset (E)^*$ over the Fock space. This triple allows not only 
the effective construction of the decomposition (along the lines presented in \cite{GelfandIV}, Chap. IV.4) 
of the restriction of the Krein-isometric representation to $SL(2,\mathbb{C})$, 
acting in the Fock space of the free e.m. potential field (whenever it is decomposable), but also, based on this decomposition, construction
of the infrared limit of the field together with the generalized states of the infrared photons. More generally, the adiabatic limit of the first order 
contribution to interacting e.m. potential, understood as integral kernel operator, admits decomposition induced by the said decomposition of
the action of $SL(2,\mathbb{C})$, and then, computation of the IR quasiasymtotics of the interacting e.m. potential, giving a concrete realization for the 
general quantum theory of the Coulomb field of \cite{Staruszkiewicz}.
 This would be impossible without $(E)$. Wightman's operator valued distribution
is not sufficient for this construction.

Finally, we should mention still another approach practiced when 
working with perturbative QFT with massless Yang-Mills fields. 
In the practical implementation of QCD (with massless Yang-Mills fields, unbroken symmetry) 
the IR-problem is treated with the help of distribution functions replacing the asymptotic states. 
Up to the author's knowledge, it is not clear at present if this method can somehow be used to save the 
scattering operator and interacting fields in the adiabatic limit, but if yes, then again, we would get a theory in which 
passing to broken phase could be avoided, and the conclusions of Section \ref{Perspectives} wolud not be true 
within this approach.

\section*{Acknowledgements}

The author would like to express his deep gratitude to Professor D. Kazakov  and Professor I. Volovich 
for the very helpful discussions.
He also would like to thank for the excellent conditions for work at JINR, Dubna.
He would like to thank Professor M. Je\.zabek 
for the excellent conditions for work at INP PAS in Krak\'ow, Poland
and would like to thank Professor A. Staruszkiewicz and 
Professor M. Je\.zabek for the warm encouragement. The author would like to acknowledge the Referees for their
suggestions.

\vspace*{1cm}

{\small Conflict of Interest: The authors declare that they have no
conflicts of interest.}


\begin{thebibliography}{}
\bibitem{Bogoliubov_Shirkov} Bogoliubov, N. N., Shirkov, D. V.: Introduction to the Theory of Quantized Fields. New York (1959), second ed. John Wiley \& Sons, Inc., New York, Chichester, Brisbane, Toronto, 1980.
\bibitem{Berezin}  Berezin, F. A.: The method of second quantization. Acad. Press, New York, London, 1966.
\bibitem{Epstein-Glaser} Epstein, H., Glaser, V.: Ann. Inst. H. Poincar\'e {\bf A19}, 211-295 (1973).
\bibitem{DKS1} D\"utsch, M., Krahe, F., Scharf, G.: Nuovo Cimento {\bf A 103}, 871 (1990).
\bibitem{DKS2} D\"utsch, M., Krahe, F., Scharf, G.: Nuovo Cimento {\bf A 1029}, 871 (1993).
\bibitem{DKS3} D\"utsch, M., Krahe, F., Scharf, G.: Nuovo Cimento {\bf A 107}, 375 (1994).
\bibitem{DKS4} D\"utsch, M., Krahe, F., Scharf, G.: Nuovo Cimento {\bf A 108}, 737 (1995).
\bibitem{Grigore}   Grigore, D. R.: J. Phys. A: Math. Gen. 33 8443 (2000).
\bibitem{wig} R. F. Streater and A. S. Wightman, {\em PCT, Spin and Statistics, and All
That}, (W. A. Benjamin, Inc., New York, 1964).
\bibitem{Vladimirov1} V. S. Vladimirov, Yu. N. Drozhzhinov, B. I. Zav’yalov, Tauberian theorems for generalized
functions in a scale of regularly varying functions and functionals, dedicated to Jovan Karamata, 
{\em Publ. Inst. Math. (Beograd)} {\bf 71}, 123 (2002) (in Russian).  
\bibitem{epstein-glaser-al} Epstein, H., Glaser, V.: Contribution to the meeting on renormalization theory. C. N. R. S., Marseille, June 1971; C. E. R. N., preprint TH 1344; reprinted in: Renormalization Theory, G. Velo and A. S. Wightman (Eds.), D. Reider Publishing Company, Dordrecht-Holland 1976, pp. 193-254.
\bibitem{obataJFA} Obata, N.: Operator calculus on vector-valued white noise functionals.  J. of Funct. Anal. 121, 185-232 (1994).
\bibitem{WN} Wawrzycki, J: Causal Perturbative QFT and white noise. To appear in:
Infinite Dimensional Analysis, Quantum Probability and Related Topics. ArXiv: math-ph $\slash$ 220305884.
\bibitem{IF} Wawrzycki, J: Theoretical and Mathematical Physics {\bf 211}, 775 (2022); ERRATUM: {\bf 212}, 1312 (2022). 
\bibitem{GelfandIV} 
I. M. Gelfand and N. Ya. Vilenkin: Applications of Harmonic Analysis: Generalized functions. Vol. 4. 
Acad. Press, New York, 1964.
\bibitem{GelfandII} Gelfand, I. M., Shilov, G. E.: Generalized Functions. Vol II. Academic Press, New York, San Francisco, London, 1968.
\bibitem{SlavnovFaddeev} Slavnov, A. A., Faddeev, L. D.: Gauge Fields, Introduction to Quantum Theory, 3-rd Russ. Ed. (2017).
\bibitem{BlaSen}  Blanchard, P., Seneor, R.: Annales de L' I. H. P. A23, 147 (1975). 
\bibitem{PDuch} Duch, P.: Ann. Inst. H. Poincar\'e {\bf 19}, 875 (2018).
\bibitem{Staruszkiewicz} Staruszkiewicz, A.: Ann. Phys. (N.Y.) 190, 354 (1989).
\bibitem{Woronowicz} Woronowicz, S. L.: Studia Mathematica, 39, 217, (1971).
\bibitem{FaddeevKulish} Kulish, P. P., Faddeev, L. D.: Theoretical and Mathematical Physics {\bf 4}, 153 (1970).
\bibitem{Scharf} Scharf, G.: Finite Quantum electrodynamics, Dover Publications, Mineola, New York, 2014.








\end{thebibliography}
\end{document}